\documentclass[twoside]{article}
\makeatletter\if@twocolumn\PassOptionsToPackage{switch}{lineno}\else\fi\makeatother

\usepackage[symbol]{footmisc}

\usepackage{tabulary,graphicx,times,caption,fancyhdr,amsfonts,amssymb,amsbsy,latexsym,amsmath}
\usepackage[utf8]{inputenc}
\usepackage[style=ieee]{biblatex} 
\bibliography{references.bib} 
\usepackage{biblatex}
\usepackage{enumitem}

\usepackage{reledmac}
\arrangementX[A]{threecol} 
\let\footnote\footnoteA
\usepackage[bookmarks=false]{hyperref}

\usepackage{url,multirow,morefloats,floatflt,cancel,tfrupee}
\makeatletter

\AtBeginDocument{\@ifpackageloaded{textcomp}{}{\usepackage{textcomp}}}
\makeatother
\usepackage{colortbl}
\usepackage{xcolor}
\usepackage{pifont}
\usepackage[nointegrals]{wasysym}
\makeatletter

\usepackage{caption}
\usepackage{subcaption}

\def\mcWidth#1{\csname TY@F#1\endcsname+\tabcolsep}

\def\cAlignHack{\rightskip\@flushglue\leftskip\@flushglue\parindent\z@\parfillskip\z@skip}
\def\rAlignHack{\rightskip\z@skip\leftskip\@flushglue \parindent\z@\parfillskip\z@skip}

\@ifundefined{etal}{}{}

\usepackage{ifxetex}
\ifxetex\else\if@twocolumn\@ifpackageloaded{stfloats}{}{\usepackage{dblfloatfix}}\fi\fi

\AtBeginDocument{
\expandafter\ifx\csname eqalign\endcsname\relax
\def\eqalign#1{\null\vcenter{\def\\{\cr}\openup\jot\m@th
  \ialign{\strut$\displaystyle{##}$\hfil&$\displaystyle{{}##}$\hfil
      \crcr#1\crcr}}\,}
\fi
}

\AtBeginDocument{%
  \@ifpackageloaded{endfloat}%
   {\renewcommand\efloat@iwrite[1]{\immediate\expandafter\protected@write\csname efloat@post#1\endcsname{}}}{\newif\ifefloat@tables}%
}%

\def\BreakURLText#1{\@tfor\brk@tempa:=#1\do{\brk@tempa\hskip0pt}}
\let\lt=<
\let\gt=>
\def\processVert{\ifmmode|\else\textbar\fi}

\@ifundefined{subparagraph}{
\def\subparagraph{\@startsection{paragraph}{5}{2\parindent}{0ex plus 0.1ex minus 0.1ex}%
{0ex}{\normalfont\small\itshape}}%
}{}

\newcommand\role[1]{\unskip}
\newcommand\aucollab[1]{\unskip}
  
\@ifundefined{tsGraphicsScaleX}{\gdef\tsGraphicsScaleX{1}}{}
\@ifundefined{tsGraphicsScaleY}{\gdef\tsGraphicsScaleY{.9}}{}
\def\checkGraphicsWidth{\ifdim\Gin@nat@width>\linewidth
	\tsGraphicsScaleX\linewidth\else\Gin@nat@width\fi}

\def\checkGraphicsHeight{\ifdim\Gin@nat@height>.9\textheight
	\tsGraphicsScaleY\textheight\else\Gin@nat@height\fi}

\def\fixFloatSize#1{}
\let\ts@includegraphics\includegraphics

\def\inlinegraphic[#1]#2{{\edef\@tempa{#1}\edef\baseline@shift{\ifx\@tempa\@empty0\else#1\fi}\edef\tempZ{\the\numexpr(\numexpr(\baseline@shift*\f@size/100))}\protect\raisebox{\tempZ pt}{\ts@includegraphics{#2}}}}

\AtBeginDocument{\def\includegraphics{\@ifnextchar[{\ts@includegraphics}{\ts@includegraphics[width=\checkGraphicsWidth,height=\checkGraphicsHeight,keepaspectratio]}}}

\DeclareMathAlphabet{\mathpzc}{OT1}{pzc}{m}{it}

\def\URL#1#2{\@ifundefined{href}{#2}{\href{#1}{#2}}}

\def\UrlOrds{\do\*\do\-\do\~\do\'\do\"\do\-}%
\g@addto@macro{\UrlBreaks}{\UrlOrds}

\edef\fntEncoding{\f@encoding}

\makeatother

\newif\ifmultipleabstract\multipleabstractfalse%
\usepackage[paperheight=11in,paperwidth=8.3in,margin=2.5cm,headsep=.7cm,top=2.5cm]{geometry}
\usepackage[T1]{fontenc}

\renewenvironment{abstract}
{\vspace*{-1pc}\trivlist\item[]\leftskip\hindawiIndent\par\vskip4pt\noindent\textbf{\abstractname}\mbox{\null}\\}{\par\noindent\endtrivlist}

 \date{} \emergencystretch 8pt

\makeatletter\def\hindawiIndent{0pc}
\def\author#1{\gdef\@author{\hskip-\dimexpr(\tabcolsep)\hskip\hindawiIndent\parbox{\dimexpr\textwidth-\hindawiIndent}{\raggedright\bfseries#1}}}

\def\title#1{\gdef\@title{\vspace*{-30pt}\raggedright\textbf{ \journaltitle}~\\\raggedright\bfseries\ifx\@articleType\@empty\vspace*{20pt}\else\vspace*{20pt}\@articleType\vspace*{20pt}\\\fi#1}}
\let\@articleType\@empty \def\articletype#1{\gdef\@articleType{{\normalfont\itshape#1}}}

\let\@runningHead\@empty \def\RunningHead#1{\gdef\@runningHead{{\normalfont #1}}}

\usepackage{fancyhdr}
\fancypagestyle{headings}{\fancyhf{}\fancyhead[R]{\thepage}\fancyhead[R]{Hindawi Template version: May18}\fancyfoot[C]{\thepage}}
\fancypagestyle{plain}{\fancyhf{}\fancyhead[R]{Hindawi Template version: May18}\fancyfoot[C]{\thepage}}

\makeatother

\def\journaltitle{\textit{Review Article}}

\usepackage{float}

\begin{document}

\nocite{*}

\title{Small-Scale 5G Testbeds for Network Slicing Deployment: A Systematic Review}
\author{Ali Esmaeily, Katina Kralevska\\[-3pt]\normalsize\normalfont \mbox{}\\{Department of Information Security and Communication Technology, NTNU - Norwegian University of Science and Technology, Trondheim 7491, Norway\\
Correspondence should be addressed to Ali Esmaeily; ali.esmaeily@ntnu.no}}

\maketitle 
\begin{abstract}
Developing specialized cloud-based and open-source testbeds is a practical approach to investigate network slicing functionalities in the fifth-generation (5G) mobile networks. This paper provides a comprehensive review of most of the existing cost-efficient and small-scale testbeds that partially or fully deploy network slicing. First, we present relevant software packages for the three main functional blocks of the ETSI NFV MANO framework and for emulating the access and core network domains. Second, we define primary and secondary design criteria for deploying network slicing testbeds. These design criteria are later used for comparison between the testbeds. Third, we present the state-of-the-art testbeds, including their design objectives, key technologies, network slicing deployment, and experiments. Next, we evaluate the testbeds according to the defined design criteria and present an in-depth summary table. This assessment concludes with the superiority of some of them over the rest and the most dominant software packages satisfying the ETSI NFV MANO framework. Finally, challenges, potential solutions, and future works of network slicing testbeds are discussed.
\end{abstract}

\maketitle

\section{1. Introduction}
\label{sec:introduction}
The fifth-generation (5G) and beyond networks are expected to provide various services compared to the 4G and previous generations of networks. The Quality of Service (QoS) requirements can be quite different in terms of low latency (or even extra-low latency), bandwidth, reliability, and availability. Remote surgery, autonomous driving, a massive number of sensors communicating with the network, and video streaming with extra high quality are just some of the numerous 5G services.The main concern here is that the physical infrastructure resources are limited and valuable, especially when data traffic demands from different operators increase. Therefore, efficient network sharing~\cite {9076108, 7470940} is considered as a conventional solution. Through network sharing, multiple operators can share infrastructure resources according to their agreed allocation plans. This approach can help an operator to reduce a significant amount of Capital Expenditure (CAPEX) and Operational Expenditure (OPEX).

As an evolution of network sharing, network slicing brings the flexibility and dynamicity of allocating the required and appropriate amount of physical resources to all service types mentioned above over the same physical infrastructure simultaneously. In fact, network slicing leverages the running of multiple logical networks on top of physical infrastructure. Network Functions (NFs)~\cite{7243304} are constructive operational components (physical networking devices) such as routers, firewalls, and load balancers that have specific functionalities in network infrastructure and hold distinct exterior interfaces for establishing communication between each other. An End-to-End (E2E) network slice~\cite{7926921} is a logical separated (isolated) network, created by chaining NFs, which delivers a particular network service according to QoS requirements via the underlying shared infrastructure in the (Radio) Access Network ((R)AN), Transport Network (TN), and Core Network (CN). 

Network Function Virtualization (NFV), Software Defined Networking (SDN), and Cloud computing are considered as the three enabling technologies for implementing network slicing in 5G. 
\begin{enumerate}[label={(\roman*)}]
\item \textit{NFV}~\cite{7243304} is a network architecture framework where NFs that traditionally used dedicated vendor-specific hardware, so-called Physical NFs (PNFs), are now implemented in software. There are two leading solutions towards softwarized PNFs: 1) Virtualized NFs (VNFs) deployed on virtual machines, and 2) Containerized NFs (CNFs) deployed on containers. 
These VNFs and CNFs, in turn, are then implemented in data centers or on cloud environments that run on top of general-purpose (vendor-neutral) hardware.
\item \textit{SDN}~\cite{6994333} enables programmable and dynamic network configuration by separating the Control Plane (CP) and the Data Plane (DP), where a centralized entity (controller) in the CP configures the forwarding devices in the DP. 
\item \textit{Cloud computing}~\cite{10.5555} deploys remote network resources in shared pools that can be administered over the Internet. Cloud computing is based on two principal orientations: 1) Cloud-based applications that point to relocating legacy applications, which were established on end-users' devices or on the companies' IT infrastructure, to cloud-based servers in order to deliver the applications over web browsers, and 2) Cloud-native applications, which refer to those applications that are particularly created and developed to employ the advantages of the cloud environment such as constant development, modularity, Application Programming Interface (API) integration, and scalability. 
\end{enumerate}

As mentioned, one of the 5G objectives is to implement ultra-low latency services and to serve many devices with different amounts of computing resources. Multi-access Edge Computing (MEC)~\cite{8016573} is an enhancement of cloud computing that reduces the latency in a mobile network by pushing the processing and computing tasks to the edges of the network (such as base stations) to be closer to the devices with a limited amount of resources. This yields in facilitating the operation of delay-sensitive applications in such devices. These enabling technologies bring flexibility, programmability, and efficiency, but at the cost of higher complexity in operating and managing the 5G networks. The necessity for the Management and Network Orchestration (MANO) framework~\cite{etsi2013network}, which performs efficient resource management and orchestration between all network elements in the whole architecture, is undeniable.
\begin{figure}[h]
\centering
\includegraphics[scale=0.55]{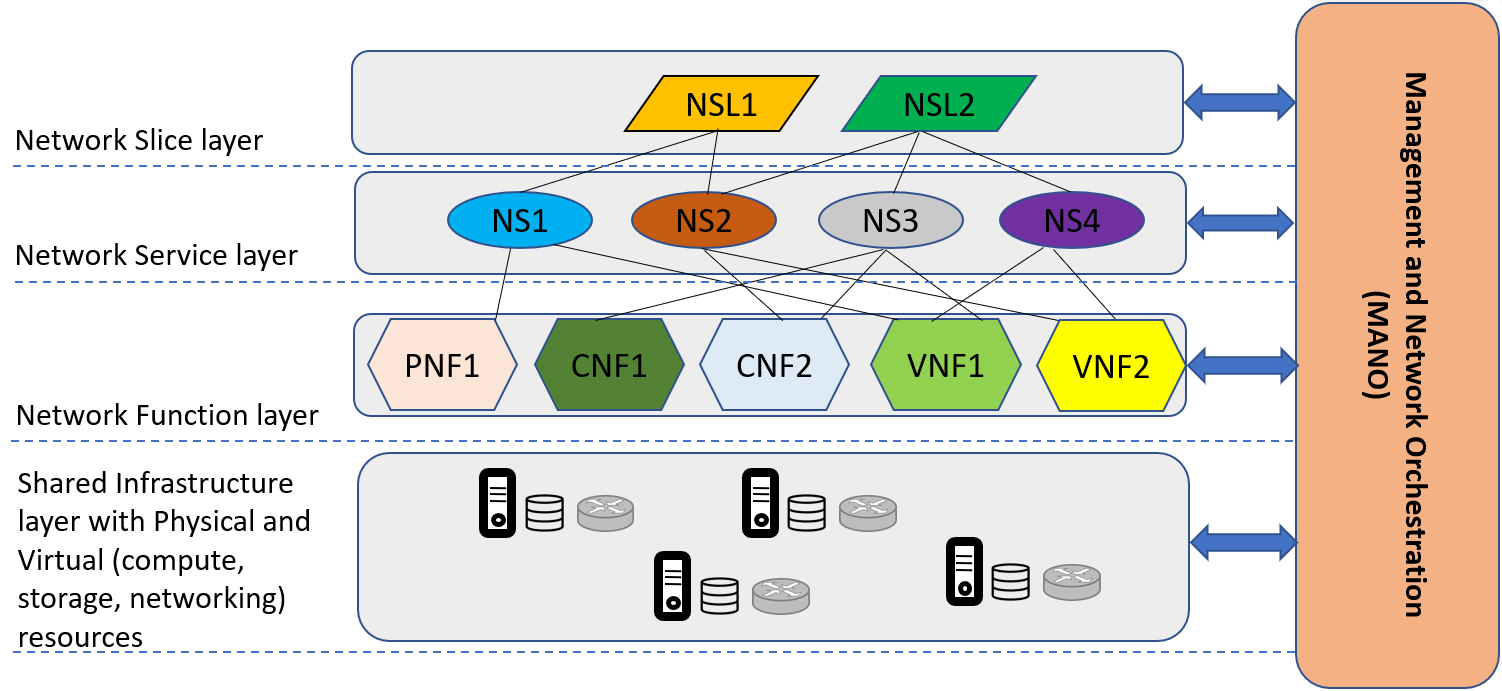}
\vspace{-0.2cm}
\caption{Multi-layered architecture of network slice provisioning in 5G via the composition of VNFs, CNF, and/or PNFs into network services to form network slices.}
\vspace{-0.1cm}\label{fig:NFV_composition}
\end{figure}

Figure \ref{fig:NFV_composition} illustrates a multi-layered architecture of network slice provisioning in 5G. 

\begin{enumerate}[label={(\roman*)}]
\item In the first layer, there is a shared infrastructure layer, which includes heterogeneous hardware and software resources (base station, compute, storage, and networking) spanning over the RAN, TN, and CN domains to host multiple NFs in the second layer. In fact, these resources are sliced according to various service requirements and then will be allocated to different service types. 
\item In the second layer, there are various NFs (PNFs, VNFs, and CNFs) with certain capabilities, belonging to different network domains. This layer encapsulates the essential configuration and managing operations of the NFs to provide different service types in the third layer.
\item In the third layer, according to service specifications, particular PNFs, VNFs, and/or CNFs (from the second layer) are chained in an explicit order with the appropriate amount of resources (from the first layer) to grant a distinctive service instance. The uniqueness of a service instance in this layer has a straightforward association with the business model, which indicates the reason for creating such a service that will be presented via a slice.
\item In the fourth layer, the launched service instances from the previous layer constitute E2E network slices. Hence, controlling and management policies on each of the network slices can be achieved independently via the NFV MANO framework.
\item The NFV MANO framework is in charge of orchestrating all of the mentioned layers. Basically, the NFV MANO delivers all the monitoring, coordinating, controlling, and managing tasks of the available physical and virtual resources in order to maintain an efficient resource utilization between all types of NFs (PNF, VNF, and CNF) in the whole architecture. This results in producing network services that meet the specific service requirements over distinctive network slices.
\end{enumerate}

Since the introduction of the network slicing concepts and specification by the 3rd Generation Partnership Project (3GPP)~\cite{3gpp2017}, network slicing has attracted a lot of attention in the past years. Apart from the theoretical aspects of different ways of achieving the 5G objectives, research communities in academia and industry have followed practical approaches to examine different features of 5G and to evaluate the network performance under various use cases. In this regard, practical research works in the 5G area have developed prototype system implementations of individual parts of the mobile network architecture, which are known as research testbeds. Recently, even more complex network architectures have been deployed on such testbeds that support network slicing. Research testbeds grant the possibility to evaluate, and enhance network performance. Besides, while research testbeds keep the cost of network deployment low, their functionalities, with a fair approximation, are comparable to real networks. Such testbeds can usually be implemented on standard PCs or servers with a not very high amount of resources and without the need of purchasing specialized hardware and software. Moreover, the availability of open-source software packages provides opportunities for creating innovative solutions towards 5G~\cite{KALTENBERGER2020107284}.

Deploying testbeds with network slicing capabilities is a challenging and error-prone task as it involves development of a network equipped with fundamental enabling technologies and the ability of programming and configuring the physical infrastructure. Depending on the specific service requirements, the physical and virtual components of a network slicing testbed must satisfy performance requests such as the amount of hardware and software resources (CPU, memory), reliability and failure rates (dependability analysis)~\cite{Stackify-website}. Nevertheless, the complexity of the testbed deployment process sometimes impacts the utilization of open-source solutions and standard PCs. 

Although some excellent surveys have been done on different aspects of network slicing such as~\cite{7926921, 8320765, 7926923}, just a few works focus on network slicing implementations; in particular,~\cite{7747513, 9003208, bonati2020open, BARAKABITZE2020106984} elaborate collaborative 5G network slicing research projects and the proposed large-scale testbeds as outcomes of these projects. Reference~\cite{7747513} presents a broad study of five main large-scale SDN testbeds by explaining their design purposes, essential technologies, slicing capability, and use cases. Reference~\cite{9003208} investigates the necessity of network slicing for facilitating the implementation of Internet-of-Things (IoT) intelligent applications and smart services. Bonati et al. in~\cite{bonati2020open} describe open source utilities, frameworks, and hardware components that can be used to instantiate softwarized 5G networks. Barakabitze et al.~\cite{BARAKABITZE2020106984} provide a comprehensive review of 5G networks, a tutorial of the 5G network slicing technology enablers including SDN, NFV, MEC, Cloud/Fog computing, network hypervisors, virtual machines, and containers, as well as an overview of collaborative large 5G network slicing implementations. Nonetheless, there is a lack of a comprehensive survey that presents and evaluates small-scale state-of-the-art 5G network slicing implementations. Small-scale network slicing testbeds are important for the research community in several aspects. Small-scale testbeds require a lower deployment budget compared to large-scale testbeds. Besides, small-scale testbeds, with a compact softwarized version of the required entities, are more effortless to deploy and launch than large-scale ones. Further, due to such testbeds' small scaling, they are more manageable to troubleshoot, and resolving possible issues is faster than large-scale testbeds with multiple involved entities. Eventually, although the number the practical use cases that can be investigated on small-scale testbeds is lower than large-scale testbeds and real networks, small-scale testbeds can afford similar analogous results to large-scale solutions. The aforementioned aspects motivate the work in this paper.  
We summarize our contributions as follows:
\begin{enumerate}[label={(\roman*)}]
\item We present the software packages and platforms that fit in the ETSI NFV MANO framework functional blocks for emulating RAN, CN domains, and MANO.
\item We define primary and secondary design criteria for network slicing testbeds.
\item We provide a detailed study of small-scale state-of-the-art testbeds for deploying network slicing. These testbeds are relatively easy to deploy and usually without requiring a huge financial investment; thus, suitable for university labs.
\item We further evaluate the testbeds according to the defined primary and secondary design criteria.
\item We highlight the typical challenges while deploying such testbeds, and present possible solutions and directions for future work.
\end{enumerate}

The rest of the paper is organized as follows. Section 2 explains the research methodology for this paper. Section 3, firstly presents the ETSI NFV MANO framework along with possible open-source software solutions for each specific block in this framework, and secondly, outlines the desired criteria for designing network slicing testbeds in 5G. In section 4, small-scale and cost-efficient state-of-the-art network slicing testbeds are detailed with their specific features. In section 5, first, we compare the testbeds presented in the previous section, and then we explain some of the main challenges while deploying such testbeds. Section 6 concludes the paper. Table \ref{TableAbbList} presents a list of the acronyms used in this paper.
\begin{table}[p!]
\vspace{-0.1cm}
\caption{List of the used acronyms in this paper.}
\vspace{-0.4cm}
\begin{center}

\begin{minipage}{\textwidth} \centering

\begin{tabular}{| m{1cm} | m{3.5cm}| m{1cm}| m{3.5cm} | m{1cm}| m{3.5cm}|}
  \hline
  \textbf{\scriptsize Abb.} & \textbf{\scriptsize Definition} & \textbf{\scriptsize Abb.} & \textbf{\scriptsize Definition} & \textbf{\scriptsize Abb.} & \textbf{\scriptsize Definition}\\ 
  \hline
    \scriptsize 5G & \scriptsize fifth-generation  & \scriptsize 4G & \scriptsize forth-generation & \scriptsize 3GPP & \scriptsize 3rd Generation Partnership Project\\ 
  \hline
  \scriptsize AI & \scriptsize Artificial Intelligence & \scriptsize AMF & \scriptsize Access and Mobility Management Function & \scriptsize API & \scriptsize Application Programming Interface\\ 
  \hline
      \scriptsize BBU & \scriptsize Baseband Unit & \scriptsize CAI & \scriptsize Connected AI & \scriptsize CN & \scriptsize Core Network\\ 
  \hline
      \scriptsize CNF & \scriptsize Containerized NF & \scriptsize CP & \scriptsize Control Plane & \scriptsize CAPEX & \scriptsize Capital Expenditure\\ 
  \hline
      \scriptsize C-RAN & \scriptsize Cloud-RAN & \scriptsize DC & \scriptsize Data Center & \scriptsize DCAE & \scriptsize Data Collection Analytics \& Events\\ 
  \hline
      \scriptsize DP & \scriptsize Data Plane & \scriptsize DSAF & \scriptsize Dynamic Slice Allocation Framework & \scriptsize E2E & \scriptsize End-to-End\\ 
  \hline
      \scriptsize eMBB & \scriptsize enhanced Mobile Broadband & \scriptsize EPC & \scriptsize Evolved Packet Core & \scriptsize ETSI & \scriptsize European Telecommunications Standards Institute\\ 
  \hline
      \scriptsize FCFSFA & \scriptsize First Come First Serve First Available & \scriptsize GUI & \scriptsize Graphical UI & \scriptsize HP LCVNF & \scriptsize High Priority LCVNF\\ 
  \hline
      \scriptsize IaaS & \scriptsize Infrastructure-as-a-Service & \scriptsize IIoT & \scriptsize Industrial IoT & \scriptsize IMS & \scriptsize IP Multimedia System\\ 
  \hline
      \scriptsize IoT & \scriptsize Internet-of-Things & \scriptsize KPI & \scriptsize Key Performance Indicator & \scriptsize KQI & \scriptsize Key Quality Indicators\\ 
  \hline
      \scriptsize L2TP & \scriptsize Layer-2 Tunneling Protocol & \scriptsize LCVNF & \scriptsize Latency Critical VNF & \scriptsize LP LCVNF & \scriptsize Low Piority LCVNF\\ 
  \hline
      \scriptsize LTE & \scriptsize Long Term Evolution & \scriptsize LT VNF & \scriptsize Latency Tolerant VNF & \scriptsize MAC & \scriptsize Medium Access Control\\ 
  \hline
      \scriptsize MANO & \scriptsize Management and Network Orchestration & \scriptsize M-CORD & \scriptsize Mobile-Central Office Re-Architected as Datacenter & \scriptsize MEC & \scriptsize Multi-access Edge Computing\\ 
  \hline
      \scriptsize ML & \scriptsize Machine Learning & \scriptsize MME & \scriptsize Mobility Management Entity & \scriptsize MTC & \scriptsize Machine Type Communication\\ 
  \hline
      \scriptsize NAS & \scriptsize Network Attached Storage & \scriptsize NBI & \scriptsize Northbound Interface & \scriptsize NF & \scriptsize Network Function\\ 
  \hline
      \scriptsize NR & \scriptsize New Radio & \scriptsize NFV & \scriptsize Network Function Virtualization & \scriptsize NFVI & \scriptsize NFV Infrastructure\\ 
  \hline
      \scriptsize NFVO & \scriptsize NFV Orchestrator & \scriptsize NIM & \scriptsize Network Infrastructure Manager & \scriptsize NSO & \scriptsize Network Service Orchestrator\\ 
  \hline
      \scriptsize OAI & \scriptsize OpenAirInterface & \scriptsize ODL & \scriptsize OpenDayLight & \scriptsize ODTN & \scriptsize Open and Disaggregated Transport Network\\ 
  \hline
      \scriptsize OMEC & \scriptsize Open Mobile Evolved Core & \scriptsize ONAP & \scriptsize Open Networking Automation Platform & \scriptsize ONOS & \scriptsize Open Network Operating System\\ 
  \hline
      \scriptsize OPEX & \scriptsize Operational Expenditure & \scriptsize OSM & \scriptsize Open Source MANO & \scriptsize OTG & \scriptsize OAI Traffic Generator\\ 
  \hline
      \scriptsize OvS & \scriptsize Open virtualSwitch & \scriptsize PaaS & \scriptsize Platform-as-a-Service & \scriptsize PNF & \scriptsize Physical NF\\ 
  \hline
      \scriptsize QoE & \scriptsize Quality of Experience & \scriptsize QoS & \scriptsize Quality of Service & \scriptsize RAN & \scriptsize Radio Access Network\\ 
  \hline
      \scriptsize RAT & \scriptsize Radio Access Technology & \scriptsize RLC & \scriptsize Radio Link Control & \scriptsize RO & \scriptsize Resource Orchestrator\\ 
  \hline
      \scriptsize RRC & \scriptsize Radio Resource Control & \scriptsize RRH & \scriptsize Remote Radio Head & \scriptsize RRM & \scriptsize Radio Resource Management\\ 
  \hline
      \scriptsize SA & \scriptsize Service Assurance & \scriptsize SaaS & \scriptsize Software-as-a-Service & \scriptsize SBI & \scriptsize Southbound Interface\\ 
  \hline
      \scriptsize SDN & \scriptsize Software Defined Networking & \scriptsize SD-RAN & \scriptsize Software Defined RAN & \scriptsize SEMIoTICS & \scriptsize Smart End-to-end Massive IoT Interoperability, Connectivity, and Security\\ 
  \hline
      \scriptsize SLA & \scriptsize Service Level Agreement & \scriptsize SlaaS & \scriptsize Slice-as-a-Service & \scriptsize SRS LTE & \scriptsize Software Radio Systems LTE\\ 
  \hline
      \scriptsize TN & \scriptsize Transport Network & \scriptsize UE & \scriptsize User Equipment & \scriptsize UI & \scriptsize User Interface\\ 
  \hline
      \scriptsize VDU & \scriptsize Virtual Deployment Unit & \scriptsize VES & \scriptsize Virtual Event Streaming & \scriptsize VIM & \scriptsize Virtualized Infrastructure Manager\\ 
  \hline
      \scriptsize VNF & \scriptsize Virtualized NF & \scriptsize VNFFG & \scriptsize VNF Forwarding Graph & \scriptsize VNFM & \scriptsize VNF Manager\\ 
  \hline
  
\end{tabular}

\end{minipage}

\end{center}
\label{TableAbbList}
\vspace{-0.4cm}
\end{table}
\section{2. Research methodology}\label{sec:research methodology}

Network slicing has become a very hot topic both in academia and industry. This trend has resulted in research on various aspects of network slicing in 5G and a fast-growing number of publications. It is evident that only a portion of these publications introduces implementation solutions for network slicing, i.e., network slicing testbeds. In order to review such publications, we followed a research methodology and defined the procedure to search for related publications, the inclusion and exclusion criteria, and finally, the data collection method to extract pertinent publications. Inclusion and exclusion criteria are used to filter out non-relevant collected papers. There is also an extra step for quality assessment regarding those publications that pass the inclusion criteria in the final systematization.

In the first step, we identified the databases for searching for potential relevant publications such as 1) ACM Digital Library, 2) IEEE Xplore, 3) Springer Link, 4) ScienceDirect, and 5) arXiv. Next, we started our searching process with relevant keywords to narrow down the selection area of the scientific publications into the network slicing field and, in particular, the deployment of network slicing. We employed some keywords such as \textit{<5G testbed>}, \textit{<network slicing testbed>}, \textit{<network slicing platform>}, and \textit{<network slicing framework>}. 

In the second step, we defined the inclusion criteria, for the publications resulted from the first step, as follows:
\begin{enumerate}[label={(\roman*)}]
\item Does the publication present a solution for network slicing deployment? 
\item How is the solution provided? Which software and hardware components are used?
\item Is the presented testbed cost-efficient in terms of equipment and also human resources needed for the tested deployment?
\end{enumerate}

We also defined the exclusion criteria as:
\begin{enumerate}[label={(\roman*)}]
\item A publication that introduces a large-scale testbed for network slicing, which is not possible to be implemented with a small budget.
\item A testbed, which is a result of national or international research projects, and those projects have been finished or are no longer active.
\end{enumerate}

In the third step, the publications that meet the inclusion criteria are assessed for their quality. Following questions are applied for quality assessment:
\begin{enumerate}[label={(\roman*)}]
\item Can the presented testbed be used to investigate different typical use cases in the 5G network slicing, or is the solution just an initial implementation of network slicing with limited capacity for providing few realistic scenarios?
\item Does the publication include comprehensive information for the testbed architecture and deployment? Are there any extra and complementary sources included in the publication, that could help other researchers to deploy a similar testbed or a possible future extension?
\end{enumerate}

In the end, we categorize the testbeds following the primary and secondary criteria defined in Section 3.

\section{3. ETSI NFV MANO framework and Design criteria for network slicing testbeds}\label{sec:ETSI MANO-design criteria}

\begin{figure}[h]
\centering
\includegraphics[scale=0.45]{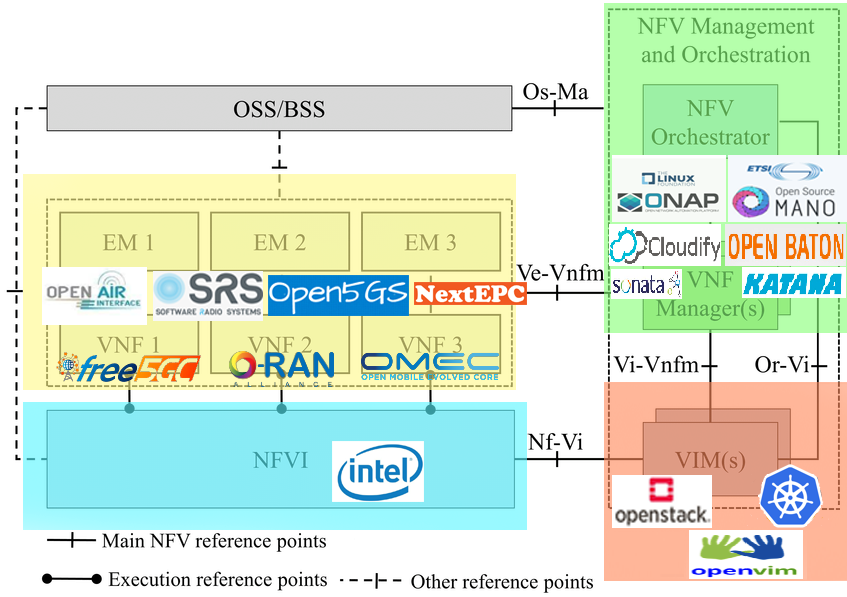}
\caption{Different open-source software solutions mapped to the ETSI NFV MANO framework~\cite{etsi2013network}.}
\label{fig:NFV_MANO}
\end{figure}

\textit{3.1. ETSI NFV MANO framework and different open-source software solutions}.
\label{sec:ETSI MANO}
ETSI introduces the NFV MANO architecture~\cite{etsi2013network}, which is comprised of three main functional blocks. Figure \ref{fig:NFV_MANO} illustrates these blocks with the reference points that connect them. This figure also summarizes some of the preeminent software solutions for each specific block. We focus on combining these solutions into the presented testbeds in Section 4 instead of explaining each one of these software modules individually. 
\begin{enumerate}[label={(\roman*)}]
\item \textit{Virtualized Infrastructure Manager (VIM)} performs controlling mechanisms for the NFV Infrastructure (NFVI) resources within an infrastructure provider. VIM is also responsible for receiving fault measurement and performance information of NFVI resources. Consequently, VIM can supervise NFVI resources allocation to the available VNFs. OpenStack~\cite{OpenStack-website} and  OpenVIM~\cite{OpenVIM-website} (for VNFs), and Kubernetes~\cite{Kubernetes-website} (for CNFs) are possible solutions for the VIM section.
\item \textit{VNF Manager (VNFM)} conducts one or several VNFs and does the lifecycle management of VNFs. VNF lifecycle management involves establishing/configuring, preserving, and terminating VNFs.
\item \textit{NFV Orchestrator (NFVO)} implements resource and service orchestration in the network. NFVO is split up into Resource Orchestrator (RO) and Network Service Orchestrator (NSO). First, RO collects the current information regarding possible physical and virtual resources of NFVI through the VIM. Second, NSO applies a complete lifecycle management of multiple network services. In this way, NFVO keeps updating the information about the available VNFs running on top of NFVI. As a result, NFVO can initiate multiple network services. As part of the lifecycle management, NFVO can also terminate a network service whenever no longer a service request is received for that specific service. In several solutions, NFVO and VNFM are integrated into the MANO section. Open Source MANO (OSM)~\cite{OSM-website}, Open Networking Automation Platform (ONAP)~\cite{ONAP-website}, OpenBaton~\cite{OpenBaton-website}, Cloudify~\cite{Cloudify-website}, SONATA~\cite{SONATA-website}, and Katana Slice Manager~\cite{Katana-website} are considered as the leading integrated solutions for the MANO section. Note that OSM can also perform management and orchestration tasks on PNFs.
\end{enumerate}

Regarding VNFs/CNFs, several open-source software solutions can emulate RAN and CN domains:
\begin{enumerate}[label={(\roman*)}]
    \item \textit{RAN} domain is emulated with Software Radio Systems LTE (srsLTE)~\cite{articlesrsLTE}, OpenAirInterface (OAI)~\cite{880944d326, KALTENBERGER2020107284}, or O-RAN in its \textit{Bronze} release~\cite{O-RAN-website, O-RAN-website-Bronze};
    \item \textit{CN} domain is realized with OAI, Open5GS (previously known as NextEPC)~\cite{Open5GS-website}, Open Mobile Evolved Core (OMEC)~\cite{OMEC-website}, or free5GC~\cite{free5GC-website}.
\end{enumerate}

Then via chaining these VNFs/CNFs in the RAN and CN by the NFVO, distinguished service instances, so-called network slice sub-instances, are formed. Some solutions for the TN domain, such as Open and Disaggregated Transport Network (ODTN)~\cite{ODTN-website}, utilize disaggregated optical equipment and open-source software to create a TN slice sub-instance. An E2E network slice instance is created by pairing the definite RAN and CN slice sub-instances via the TN slice sub-instance~\cite{8698758}.\\\\
\textit{3.2. Design criteria for network slicing testbeds}.
\label{sec:design criteria}
Multiple features should be taken into consideration when designing a comprehensive testbed of 5G and beyond networks. We identify the key design criteria for creating a 5G testbed that can emulate a real network's major features and allow us to develop and test new algorithms. They are divided into two groups.

(1) Primary criteria
\noindent

These attributes are fundamental for creating a network slicing testbed. 
\begin{enumerate}[label={(\roman*)}]
\item \textit{Support of the main enabling technologies.} The proposed testbed should be based on SDN, NFV, and cloud computing. Therefore, flexibility and dynamicity in the network are granted. SDN and NFV are complementary, hence, combined with cloud computing pave the way for the paradigms Software-as-a-Service (SaaS), Platform-as-a-Service (Paas), and Infrastructure-as-a-Service (IaaS)~\cite{7243304}.
\item \textit{MANO equipped with dynamic monitoring capability.} The testbed should support management, orchestration, programmability, and dynamic monitoring of different network functions, network services, and network slices. Therefore, the role of the MANO entity is essential that is the result of SDN/NFV utilization in the network architecture~\cite{101145}. 
\item \textit{Multi-network domain with partial slicing support.} A 5G testbed needs to provide connectivity across all network domains (air interface, (R)AN, TN, CN) in order to show a practical ability that emulates the main functionalities of the 5G network. Multi-network domain support allows achieving E2E network slicing; however, it is worth noting that network slicing is a capability that can be implemented partially, and testbeds can deploy slicing only in one specific network domain.
\item \textit{Multi-tenancy support.}
5G network is expected first to enable the co-existence of multiple tenants that demand the same network functionalities, and second to administrate the cooperation and interaction between them. This capability represents the so-called multi-tenancy environment, which means that a single instance of the software and its supporting infrastructure serves multiple tenants. Multi-tenancy is one of the main aspects of the 5G networks and should be supported in the testbed implementation.
\end{enumerate}

(2) Secondary criteria
\noindent

These attributes add extra features to a network slicing testbed apart from those in the \textit{primary} group. Testbeds with these extra features broaden the research scope in the network slicing field.
\begin{enumerate}[label={(\roman*)}]
\item \textit{Multi-radio access technologies support.} Different Radio Access Technologies (RATs) such as Long Term Evolution (LTE), WiFi, and 5G New Radio (5G NR) should be deployed on the same platform~\cite{8108594}. Furthermore, Cloud-RAN (C-RAN), as a cloud computing-based architecture, brings cloudification benefits into the RAN domain. C-RAN consists of a cloud-Baseband Unit (BBU) pool and several Remote Radio Heads (RRHs). Since the 5G-RAN domain integrates the mentioned RATs with the corresponding frequency bands and provides them via the cloud, a solid platform should implement these capabilities. 
\item \textit{End-to-End network slicing.} The slicing capability should be expanded upon all network domains. An E2E network slice consists of several network slice subnet instances, each belonging to a particular network domain. Therefore, all network slice subnet instances should be provided and chained together to form an E2E network slice.
\item \textit{Cross-location support.} One possible solution for experimenting with more realistic scenarios is deploying testbeds located in two geographical areas. In this case, RAN and CN domains are implemented and launched on two geographically separated infrastructures, and a backbone TN interconnects them. The cross-location capability becomes even more essential when evaluating network performance for providing delay-sensitive services in the 5G network. In real-world use cases, the RAN and CN domains are not necessarily located in the same geographical location, and, as mentioned, MEC is the technology answer to expedite the communication between the RAN and CN domains. Hence, cross-location testbeds facilitate measuring service delay and proposing possible solutions for those services that require low latency.
\item \textit{Machine Learning (ML)-enabled.} 5G testbeds equipped with ML toolkits enable users to design, verify, and operate machine learning models via a supervised user interface. One possible outcome of using ML techniques in network slicing is to predict wireless channel behavior in the RAN domain. As a result, the available radio resources can be scheduled in an optimized way to maximize the resource usage per end-user or slice in the next transmissions.
\item \textit{Open-source.} Providing open-source 5G platforms with well-defined interfaces is considered as a huge advantage in deploying 5G testbeds because an open-source testbed can be deployed by other researchers to help foster research and innovation. It helps to reduce the hassle of setting up a working mobile network that on itself is a complicated and error-prone process.
\end{enumerate}
\begin{figure}[h]
\centering
\includegraphics[scale=0.45]{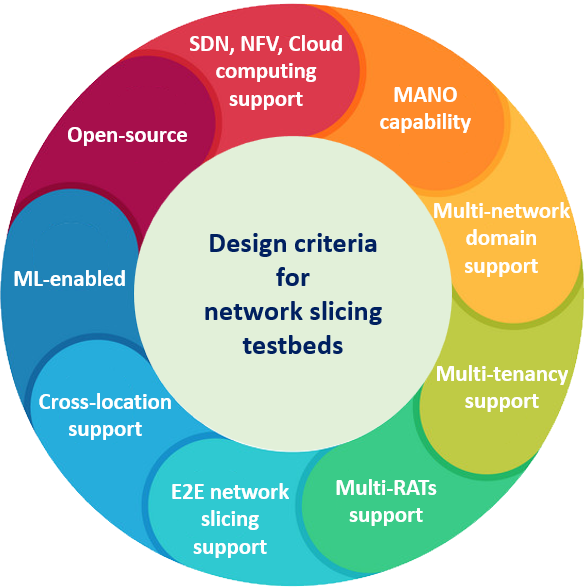}
\caption{Design criteria for network slicing testbeds.}
\label{fig:Design_criteria}
\end{figure}

These design criteria explained above and outlined in Figure \ref{fig:Design_criteria}, are later used as an assessment for the state-of-the-art testbeds (represented as columns in Table \ref{TableOverview}). 

\section{4. An overview of the state-of-the-art network slicing testbeds}
\label{sec:state-of-the-art-testbeds}
\begin{figure}[h]
\centering
\includegraphics[scale=0.51]{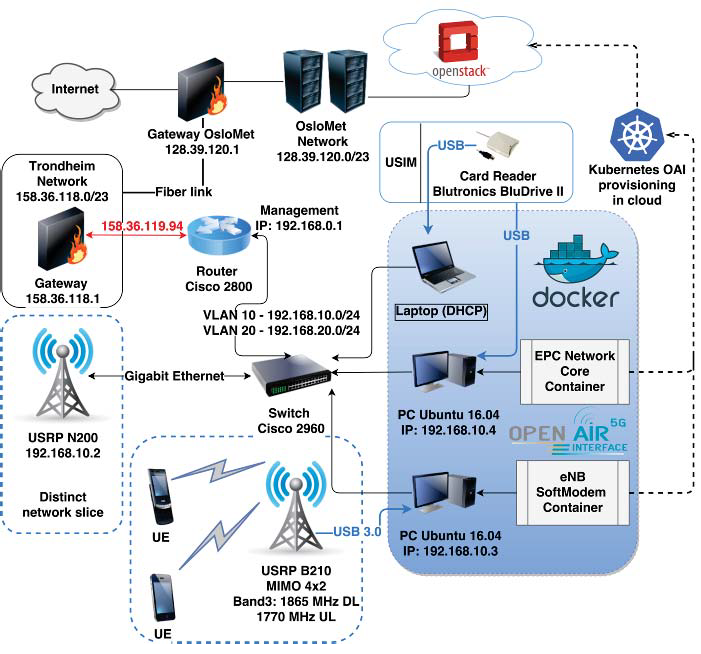}
\vspace{-0.1cm}
\caption{5G4IoT testbed architecture~\cite{8405499}.}
\vspace{-0.2cm}\label{fig:5G4IoT}
\end{figure}

\begin{figure}[h]
\centering
\includegraphics[scale=0.55]{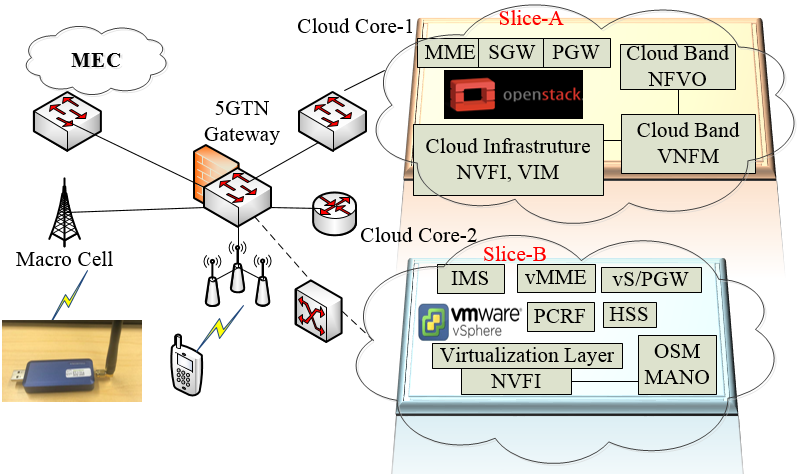}
\vspace{-0.2cm}
\caption{5GTN testbed architecture~\cite{8902745}.}
\vspace{-0.2cm}\label{fig:5GTN}
\end{figure}

\begin{figure}[h]
\centering
\includegraphics[scale=0.65]{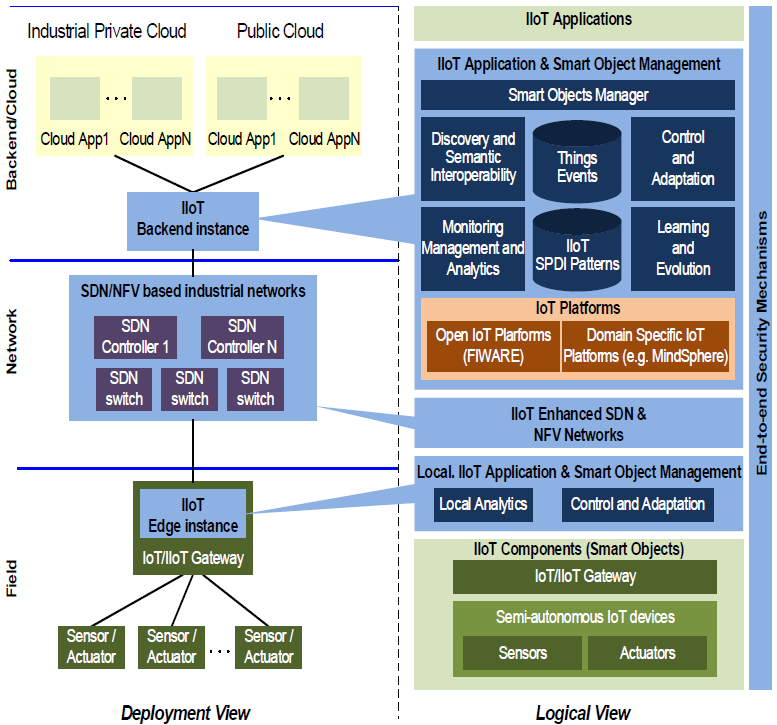}
\vspace{-0.1cm}
\caption{Architecture of SEMIoTICS platform~\cite{SEMIoTICS_framework}.}
\vspace{-0.6cm}\label{fig:SEMIoTICS}
\end{figure}

\begin{figure}[h]
\centering
\includegraphics[scale=0.59]{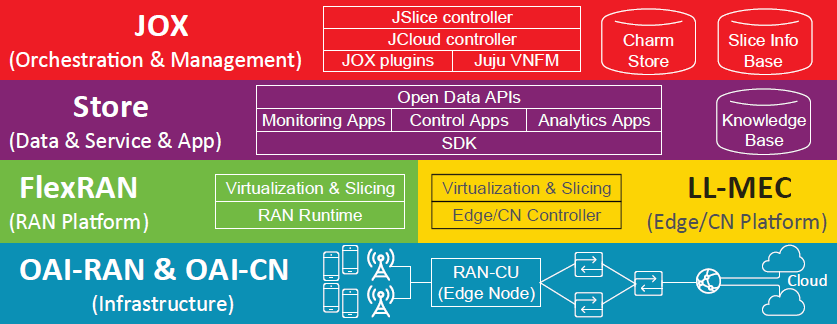}
\vspace{-0.2cm}
\caption{Mosaic5G platform schematic architecture~\cite{102623}.}
\vspace{-0.2cm}\label{fig:Mosaic5G}
\end{figure}

\begin{figure}[h]
\centering
\includegraphics[scale=0.55]{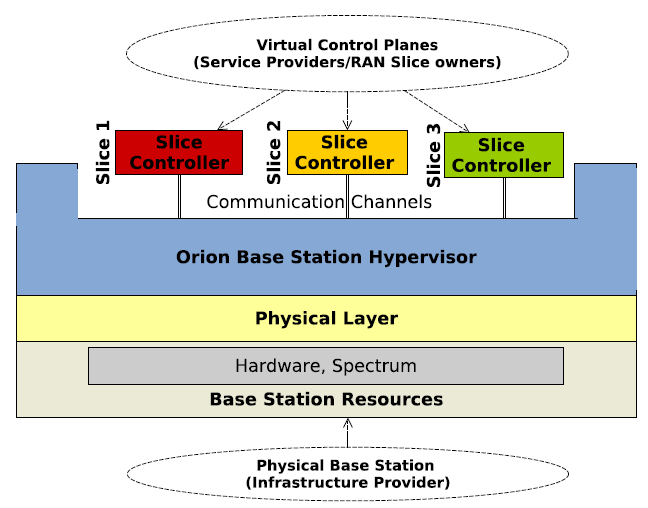}
\vspace{-0.3cm}
\caption{High level of Orion testbed architecture~\cite{Foukas2017OrionRS}.}
\vspace{-0.4cm}\label{fig:Orion}
\end{figure}

\begin{figure}[h]
\centering
\includegraphics[scale=0.51]{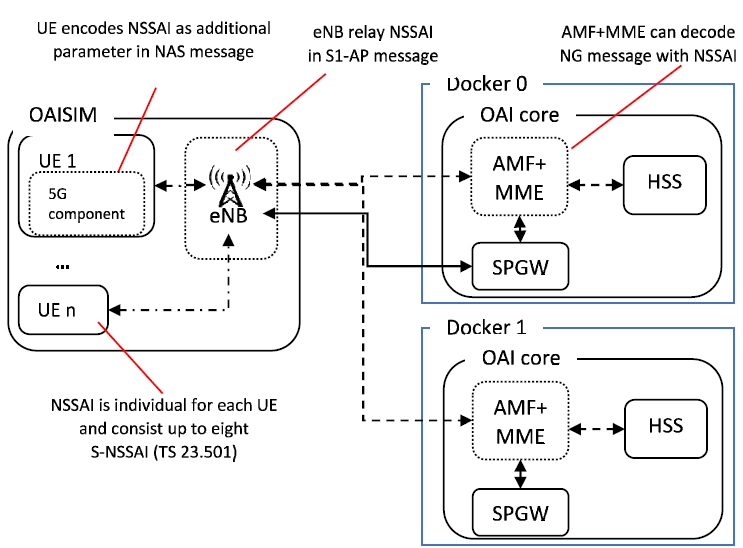}
\vspace{-0.1cm}
\caption{5G testbed with network slicing support~\cite{865686122356}.}
\vspace{-0.2cm}\label{fig:5G_Testbed_for_NS}
\end{figure}

\begin{figure}[h]
\centering
\includegraphics[scale=0.9]{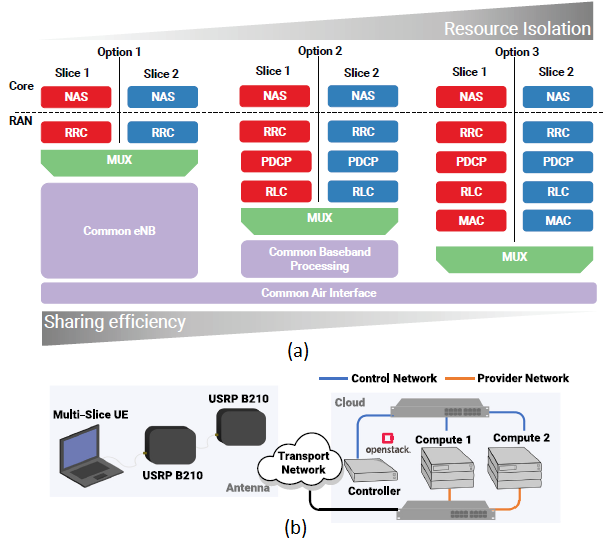}
\vspace{-0.1cm}
\caption{POSENS testbed (a). RAN slicing options. (b). Architecture~\cite{8524891, garcia2020experimenting}.}
\vspace{-0.2cm}\label{fig:POSENS}
\end{figure}

\begin{figure}[h]
\centering
\includegraphics[scale=0.47]{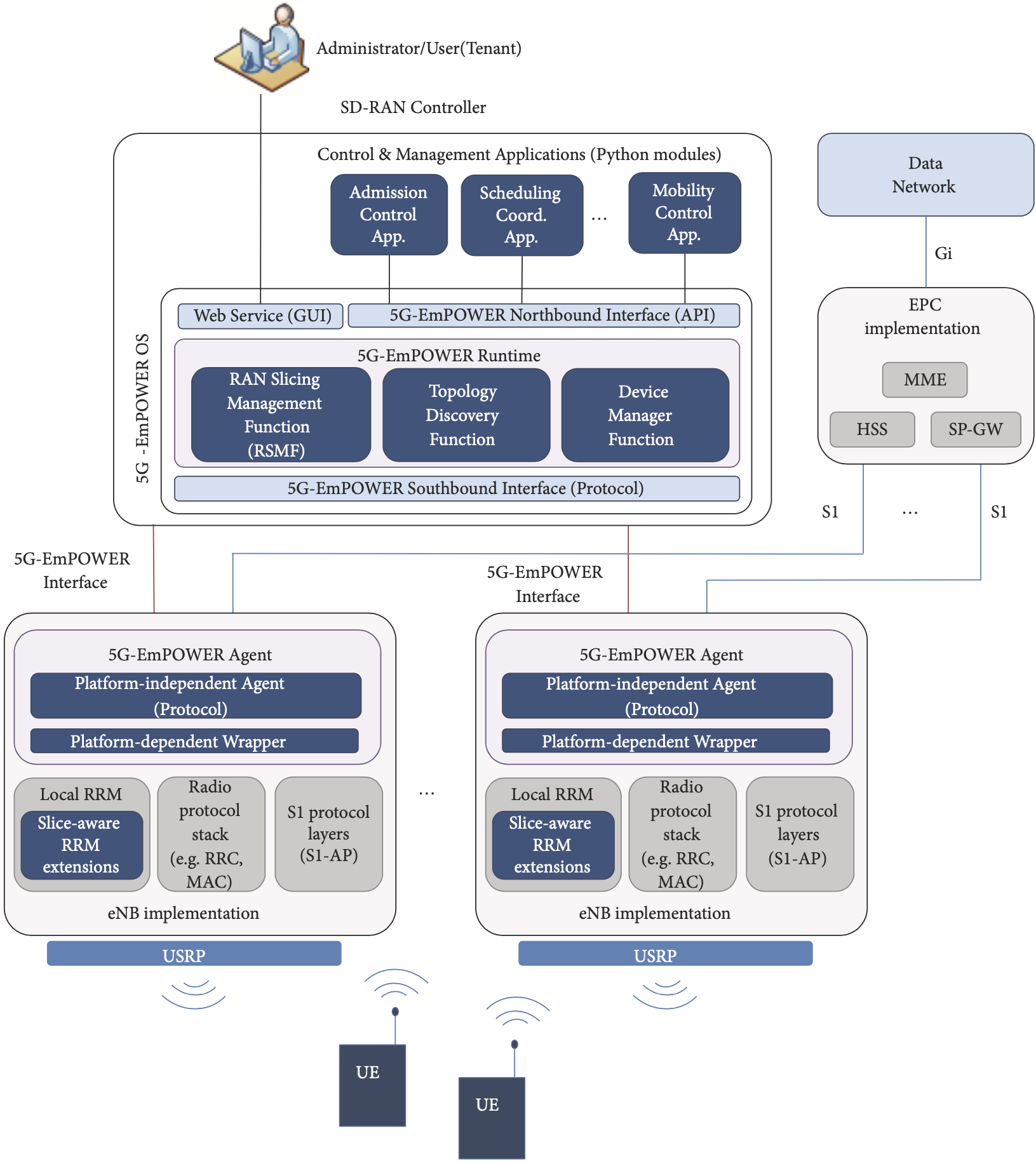}
\vspace{-0.1cm}
\caption{UPC testbed integrated with 5G-EmPOWER SD-RAN controller~\cite{articleUPC}.}
\vspace{-0.2cm}\label{fig:UPC_testbed}
\end{figure}

\begin{figure}[h]
\centering
\includegraphics[scale=0.33]{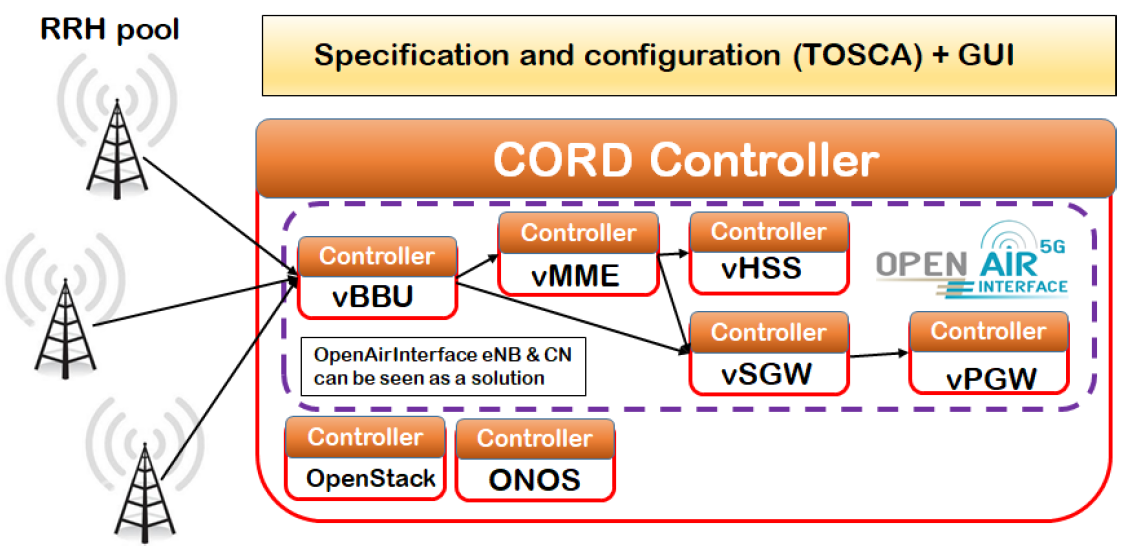}
\vspace{-0.1cm}
\caption{M-CORD framework schematic architecture ~\cite{8631816MCORD}.}
\vspace{-0.2cm}\label{fig:M_CORD_based_testbed}
\end{figure}

\begin{figure}[h]
\centering
\includegraphics[scale=0.47]{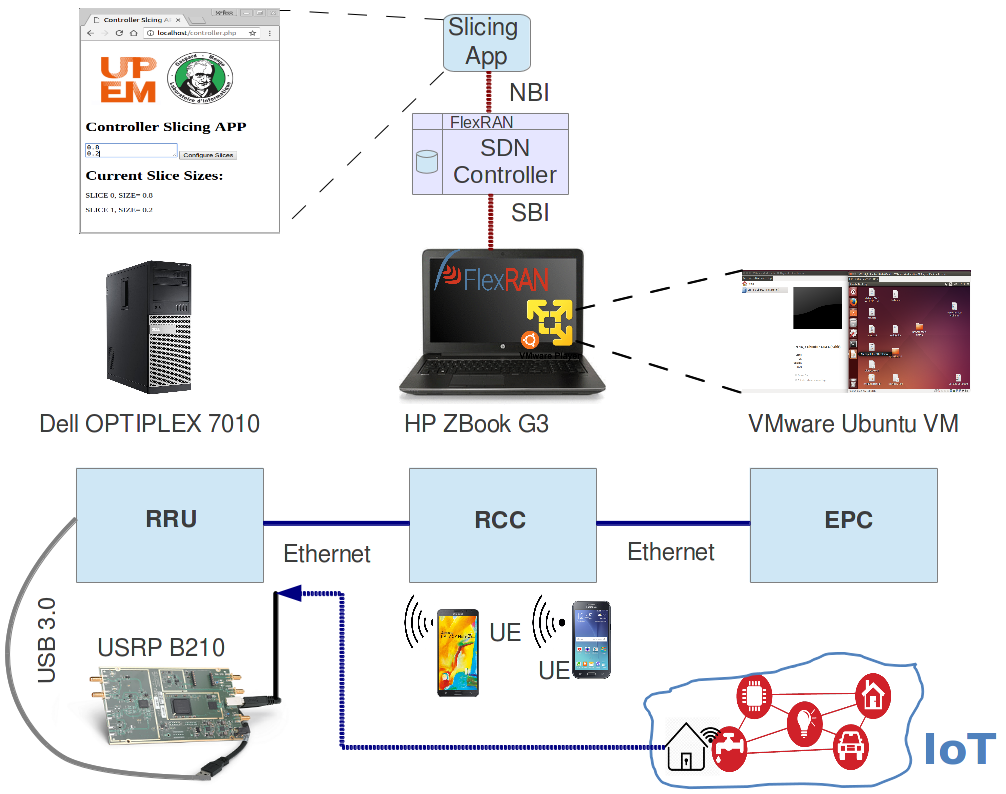}
\vspace{-0.1cm}
\caption{5G network slicing testbed supporting IoT and eMBB services~\cite{8627115}.}
\vspace{-0.2cm}\label{fig:NS_for_5G_IoT_and_eMBB}
\end{figure}

\begin{figure}[h]
\centering
\includegraphics[scale=0.31]{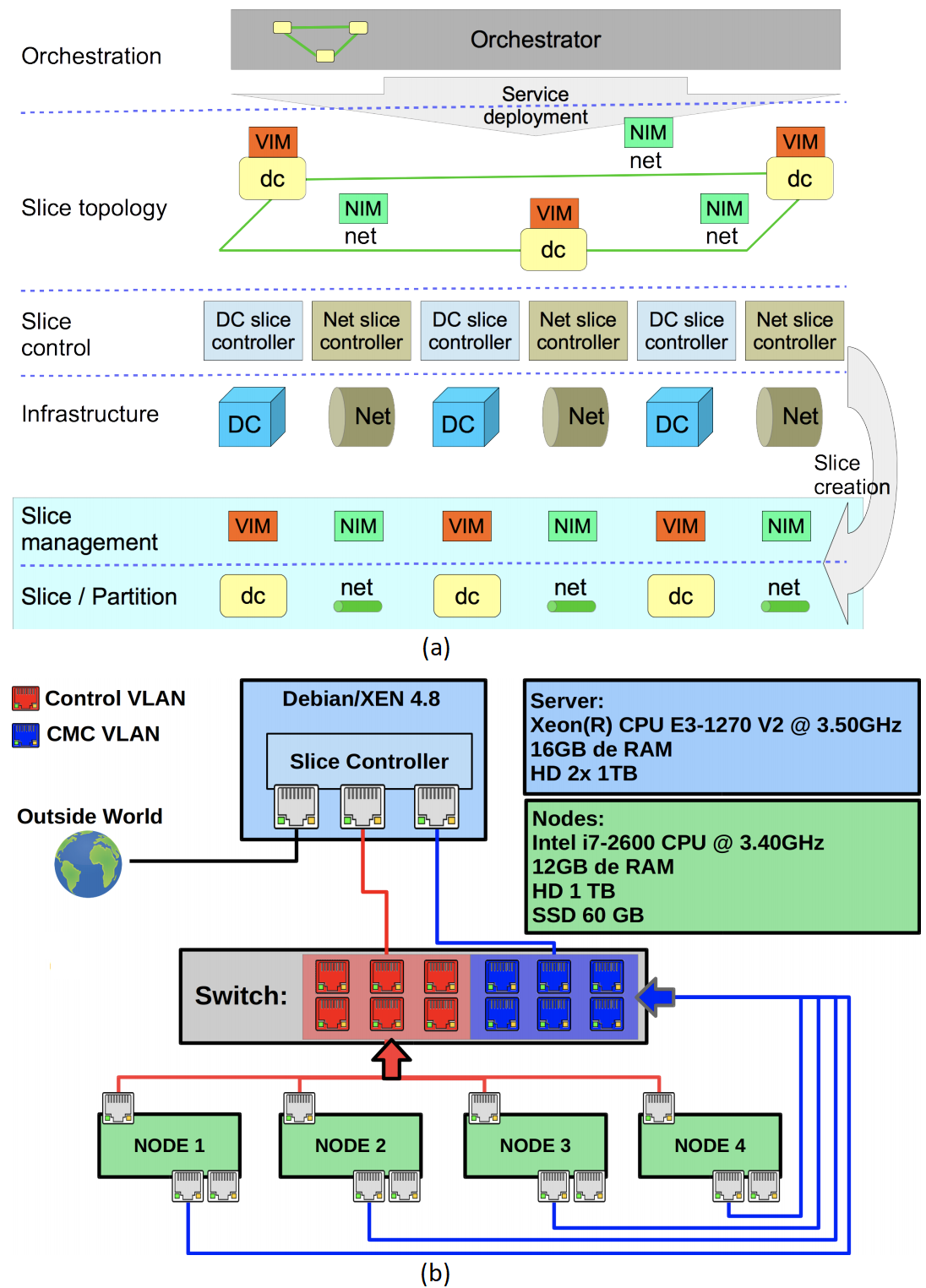}
\vspace{-0.1cm}
\caption{Transformable resources slicing testbed (a). SlaaS. (b). Architecture.~\cite{8459990}.}
\vspace{-0.2cm}\label{fig:slicing_of_transformable_resources_of_vims}
\end{figure}

We describe most of the state-of-the-art testbeds designed for implementing network slicing. We exclude large scale testbeds that are costly in terms of equipment and human resources as some of them are already explained in \cite{bonati2020open, BARAKABITZE2020106984}.\\\\
\textit{4.1. 5G4IoT~\cite{8405499, 8854015}}. This testbed (Figure \ref{fig:5G4IoT}) deploys OAI in containers to virtualize both Evolved Packet Core (EPC) and eNB. For scalability purposes, the testbed has been implemented in several containers. The testbed is created on a cloud infrastructure based on OpenStack, which is located at Oslo Metropolitan University. There are also a cisco switch and a cisco router in this testbed, separated into two VLANs, which establish the connection between EPC and eNB in two ways. The first method uses SDN Calico~\footnote{\url{https://projectcalico.org}} for layer-3 packet exchange, which provides scalability and dynamic security on the cloud infrastructure for layer-3 routing for IoT. The second approach is Layer-2 Tunneling Protocol (L2TP) to encapsulate the traffic for those IoT applications that need a lower security level. The latter approach is granted by Open virtualSwitch (OvS) without using IPSec. In the testbed, OpenStack Heat templates, as an underlay networking policy, are used to integrate OAI EPC in the OpenStack Neutron. These templates manage cloud-based applications in a stack of containers, and various services via network slices can be created. The testbed is evaluated by producing two isolated network slices for eHealth and light Internet on the same infrastructure.
\vspace{-0.2cm}
\\\\
\textit{4.2. 5G Test Network (5GTN)~\cite{8902745}}. In this testbed (Figure \ref{fig:5GTN}), located at Oulu University, the RAN operates on licensed LTE and 5G bands. The CN comprises EPC and IP Multimedia System (IMS). The CN components are implemented on cloud infrastructure, OpenStack and VMWare. The testbed aims to serve different use cases; thus, it includes MEC in the edge to provide low latency services. However, there is no RAN slicing, so only CN slicing is currently implemented. The testbed includes multiple CN domains, which result in sharing radio resources for different services. In this case, each base station in the RAN utilizes a single gateway to access a slice. The authors showcased two slices, slice A and slice B, provided in the CN domain. Slice A from the EPC domain (deployed on OpenStack and orchestrated by CloudBand which is the Nokia platform for NFV orchestration) provides enhanced Mobile Broadband (eMBB) services for IoT and content delivery applications, and slice B in the IMS domain (deployed on VMWare and orchestrated by OSM) provides critical communication and Voice over LTE services. By changing the Access Point Name between EPC (deployed on OpenStack) and IMS (deployed on VMWare), User Equipment (UE) switching between the two slices is possible. The testbed has been examined for CPU utilization, throughput, and delay for the two specific slices.
\vspace{-0.2cm}
\\\\
\textit{4.3. 5G Tactile Internet platform~\cite{8718538}}. This testbed (Figure \ref{fig:SEMIoTICS}) follows the Smart End-to-end Massive IoT Interoperability, Connectivity, and Security (SEMIoTICS)~\cite{SEMIoTICS_framework} architecture to create a 5G platform based on SDN, NFV, and MEC. The principal objective of SEMIoTICS is to build a framework to provide secure and reliable E2E services with sub-millisecond latency in actuation operations for IoT/Industrial IoT (IIoT) applications. The SEMIoTICS architecture consists of 3 layers: Backend/Cloud, Networking, and Field layers. The Backend layer is a cloud-based OpenStack platform. It creates several VMs and performs their lifecycle management. The services are provided in several containers and managed by OpenStack Tacker. Currently, there are two deployed VNFs; one for smart monitoring and one for actuating. The Networking layer manages the virtual domain on the testbed and creates tenants to chain VNFs by utilizing the SDN controller Neutron. The communication between separated tenants and also with external networks takes place by performing layer 3 routing. The Field layer is responsible for establishing a connection between smart sensors and actuators with the upper layers. This process is done by exchanging messages through IoT/IIoT gateways in the Field layer and virtual SDN switches in the Networking layer. The testbed performance has been assessed for performing E2E slicing and dynamically sharing the available bandwidth between the two VNFs.
\\\\
\textit{4.4. Mosaic5G~\cite{102623}}. Mosaic5G platform (Figure \ref{fig:Mosaic5G}) brings flexibility and scalability to service provision. The testbed architecture consists of five software modules along with hardware components: OAI, FlexRAN, LL-MEC, Store, and JOX. OAI emulates both the RAN and the CN domains of LTE networks. FlexRAN~\cite{299959999} is an open-source Software Defined RAN (SD-RAN) entity. FlexRAN delivers one of these two tasks, deployment of controlling mechanisms for several base stations in a centralized way or performing distributed controlling policies. These two actions are done as reconfigurable control functionalities in the RAN domain. LL-MEC separates the control plane and data plane traffic at the edge and the CN domain. In this way, the MEC functionality is achieved. Basically, FlexRAN and LL-MEC perform SDN functionality in the RAN, and in the edge and core domains, respectively. Store includes a set of modules, monitoring and control applications for developing network applications for a specific use case. JOX plays the role of orchestration in the network to provide several E2E network slices according to NFV MANO platform. Therefore, network slices can be deployed and then optimized based on various service specifications. The Mosaic5G platform has been used for a few use cases such as critical e-Health, V2X communication for intelligent transportation systems, and multi-service management/orchestration for smart cities.
\vspace{-0.2cm}
\\\\
\textit{4.5. Orion~\cite{Foukas2017OrionRS, foukas2018experience}}. The proposed architecture of the Orion testbed (Figure \ref{fig:Orion}) enables dynamic RAN slicing. Orion provides the sharing of RAN resources in addition to applying isolation between slices, and so, operation in one slice cannot degrade the performance of another slice. This is achieved by having an independent control plane in the RAN domain for each slice. As a result, Orion offers the opportunity to deploy different service characteristics in the RAN domain, and it is a concrete step towards realizing RAN-as-a-Service. The testbed consists of two main components: Base Station Hypervisor and virtual control plane. The Base Station Hypervisor performs slice isolation in the RAN domain, while the Hypervisor capability prepares an abstraction layer of available radio resources to the slices in the RAN. In this way, service providers build virtual base stations on top of the Hypervisor in order to create their RAN slices. A separated virtual control plane for each slice interconnects to the Hypervisor to exchange the required signaling messages. This independent deployment enables slice isolation in the RAN domain. Furthermore, the Orion architecture enables a virtual control plane of a slice to connect to multiple base stations via their Hypervisor layer. Several case studies regarding Orion's performance evaluation have been done, such as testing slice isolation and possibilities for E2E- and multi-service provisions.
\vspace{-0.2cm}
\\\\
\textit{4.6. 5G Testbed for Network Slicing Evaluation~\cite{865686122356}}. This testbed (Figure \ref{fig:5G_Testbed_for_NS}) utilizes OAI for both RAN and CN domains. There are two CNs which share radio resources of a single eNB in the RAN. OAI RAN consists of OAISIM that allows simulation of UE and eNB. OAISIM acts like a real RAN domain and simulates the LTE protocol stack. A UE with a Network Slice Selection Assistance Information (NSSAI) capability has been implemented in the testbed. Deploying two CNs in containers provides CN slice isolation. In both CNs, the Access and Mobility Management Function (AMF), which is one of the entities in 5G architecture, has been integrated with the Mobility Management Entity (MME) of the LTE platform. The eNB in the RAN selects a CN according to the NSSAI information, which is provided via the S1-AP interface between the eNB and each CN. The testbed has been appraised for connection establishment for both normal LTE UEs and UEs with an implemented NSSAI capability. In the case of normal LTE UE, it includes required encoding messages during the attach process to the network. In the case of the modified UEs, NSSAI is implemented as an optional field in them. The related CN decodes this NSSAI via the S1-AP interface provided by the eNB in the attach process to the network.
\vspace{-0.2cm}
\\\\
\textit{4.7. POSENS~\cite{8524891, garcia2020experimenting}}. POSENS platform (Figure \ref{fig:POSENS}) provides efficient resource utilization for creating independent and customizable E2E slices. Different NFs in the network layer are chained by MANO to create network slices. Then, the slices for different tenants need to be multiplexed/demultiplexed over shared resources. This procedure requires enabling the capability of multiple CNs connections to a single RAN domain and the possibility for a UE to benefit from more than one slice simultaneously. In POSENS, the possibility of implementing RAN slicing is discussed via three options: 1) Slice-aware shared RAN (slicing protocol stack down to Radio Resource Control (RRC)), where the whole radio domain is shared but CNs are distinguished by the specific services they provide and a UE can utilize different slices provided by the CNs; 2) Slice-specific radio bearer (slicing protocol stack down to Radio Link Control (RLC)), where only cell-specific functionality is shared; and 3) Slice-specific RAN (slicing protocol stack down to Medium Access Control (MAC)), which apart from the air interface, slices of different tenants are isolated in other protocol stack layers. The latter case needs specific synchronization policies among slices to be deployed efficiently. Each option holds its own level of performance and implementation complexity and POSENS implements the first option for RAN slicing in its first release. For CN, POSENS utilizes OAI CN with no modifications. In the case of MANO, the testbed provides per-slice management and an orchestration mechanism deployed in customized version of OSM. The testbed has been evaluated in terms of independency between slices, throughput, and independent service function chaining.
\vspace{-0.2cm}
\\\\          
\textit{4.8. UPC University testbed~\cite{articleUPC}}. UPC platform (Figure \ref{fig:UPC_testbed}) implements RAN slicing via RESTful API automatically. The testbed applies the slice-aware policy in Radio Resource Management (RRM) functionalities for admission control and scheduling processes. 5G-EmPOWER~\cite{8680665}, acting as a SD-RAN, allows RAN slicing management, and it also shares the available radio resources among the created RAN slices according to RRM descriptors. The interconnection between 5G-EmPOWER and eNB in the RAN is provided via an \textit{EmPOWER Agent} for performing management policies in the data plane. The testbed utilizes OAI or Next EPC for the CN domain. The srsLTE emulates LTE eNB. The \textit{EmPOWER Agent}, in turn, splits up into two sections: 1) \textit{Agent}, which includes protocol parser for EmPOWER exchanged messages and manager entities for different message types. The message type is changed depending on the requested message originated from either the 5G-EMPOWER or the \textit{Agent}, and activation/deactivation of a specific capability on the \textit{Agent} side; 2) \textit{Wrapper}, which converts EmPOWER messages to LTE protocol stack information. Several practical scenarios have been carried out for implementing RRM functionalities for admission control, scalability of the network, isolation among the existent slices.
\vspace{-0.2cm}
\\        
\textit{4.9. Mobile-Central Office Re-Architected as Datacenter (M-CORD) based 5G framework~\cite{8631816MCORD, 840611333MCORD}}. This work focuses mainly on OAI integration with the M-CORD framework (Figure \ref{fig:M_CORD_based_testbed}) and different implementation procedures to deploy LTE network on top of M-CORD. The testbed in~\cite{840611333MCORD} extends the previous work further by deploying two CN instances connecting to the C-RAN architecture via the TN in order to slice and manage the TN domain. Notably, different phases of a slice lifecycle from provisioning, allocating a slice to a UE, and managing the slice are provided by this framework. Several entities are integrated into M-CORD which emulate a complete network. XOS performs service orchestration while OpenStack provides the infrastructure for deploying the services via chaining VNFs. Open Network Operating System (ONOS)~\cite{10.1145/2620728.2620744} acts as an SDN controller and separates CP and DP functionalities. Available resources are modified and configured/reconfigured via Graphical User Interface (GUI). TN slicing is performed by running slicing policy via ONOS SDN to establish a connection flow between CN and RAN domains. In fact, ONOS inquires the OpenStack via REST API to receive the necessary information regarding the underlying platform to create TN slice between the CN and RAN domains. ONOS performs management mechanisms on TN slices via its Southbound Interface (SBI).
\vspace{-0.2cm}
\\\\
\textit{4.10. Dynamic Network Slicing for 5G IoT and eMBB services~\cite{8627115}}. This testbed (Figure \ref{fig:NS_for_5G_IoT_and_eMBB}) demonstrates sharing of the same RAN resources among eMBB and IoT services. A Northbound SDN application is designed in this testbed to create isolated RAN slices for IoT and eMBB devices according to their service requirements. IoT devices connect to the C-RAN via a gateway. The real-time slicing decision in C-RAN is performed by an SDN controller (FlexRAN) that connects via its Northbound Interface (NBI) to a Slicing app entity, which includes IoT and eMBB modules. With the help of the scheduling process conducted by the SDN controller, the slicing app determines the number of allocated radio resources to each specific slice. In the testbed architecture, the LTE scheduling mechanism is operated by the SDN controller, where CP of the MAC layer is administered as a Northbound SDN application on the cloud. An agent module is responsible for connection establishment between the slicing scheduler entity in DP and the SDN controller in CP via the SBI. Other actions, such as admission control decision, duration of allocating radio resources to a slice, are also performed by the SDN controller and the slicing app. The testbed has been evaluated in some scenarios, such as measuring average downlink throughput in IoT and eMBB slices.
\vspace{-0.2cm}
\\\\
\textit{4.11. Transformable resources slicing testbed for deployment of Multiple-VIMs~\cite{8459990}}. This testbed (Figure \ref{fig:slicing_of_transformable_resources_of_vims}) concentrates on providing Slice-as-a-Service (SlaaS) considering Data Centers (DCs). In this case, a slice is composed of a combination of DC slices (compute and storage resources) attached by Network slices (networking resources) operating on their own VIMs and Network Infrastructure Managers (NIMs), respectively. This work is considered as a supplement for Clayman's model~\cite{clayman2017network}. Clayman's model consists of three layers: 1) Orchestration layer, which manages the slice lifecycle, optimizes resource allocation, and coordinates DC slices and Network slices of a particular slice; 2) DC slice and Network slice controllers layer; former creates DC slices and deploys upon request VIM and later creates Network slices between DC slices and deploys upon request NIM; 3) Infrastructure layer, which includes all physical resources. 

\begin{figure}[h]
\centering
\includegraphics[scale=0.5]{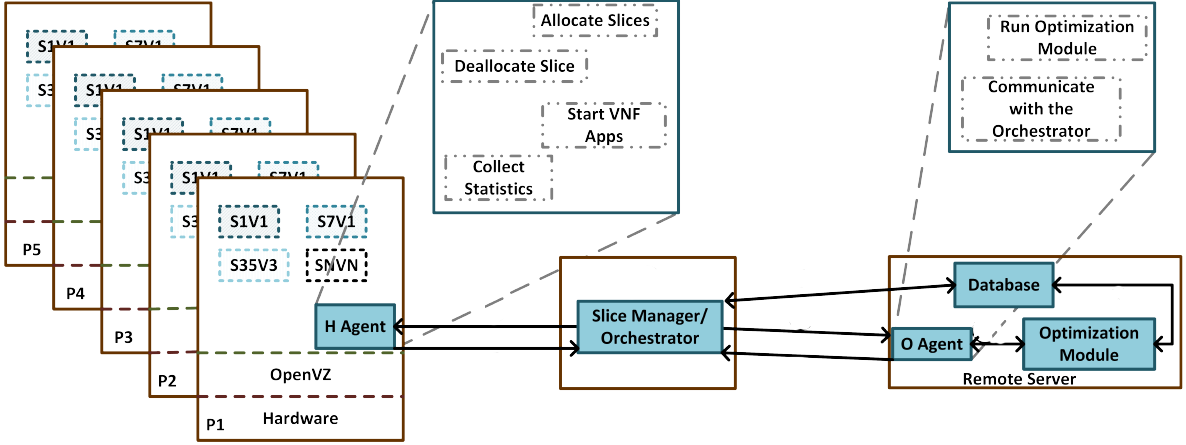}
\vspace{-0.2cm}
\caption{Implementing a topology with physical servers (brown blocks) and DSAF components (blue blocks with solid lines representing logical communication paths)~\cite{Sattar2019DSAFDS}.}
\vspace{-0.2cm}\label{fig:DSAF}
\end{figure}

\begin{figure}[h]
\centering
\includegraphics[scale=0.6]{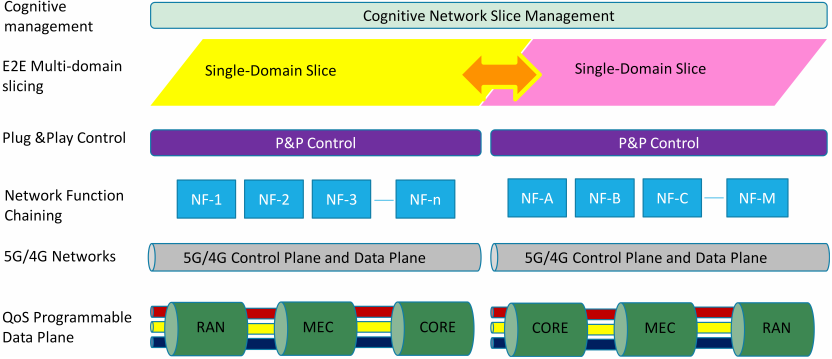}
\vspace{-0.1cm}
\caption{E2E network slicing approach in SliceNet platform~\cite{8667001}.}
\vspace{-0.2cm}\label{fig:SliceNet}
\end{figure}

\begin{figure}[h!]
\centering
\includegraphics[scale=0.75]{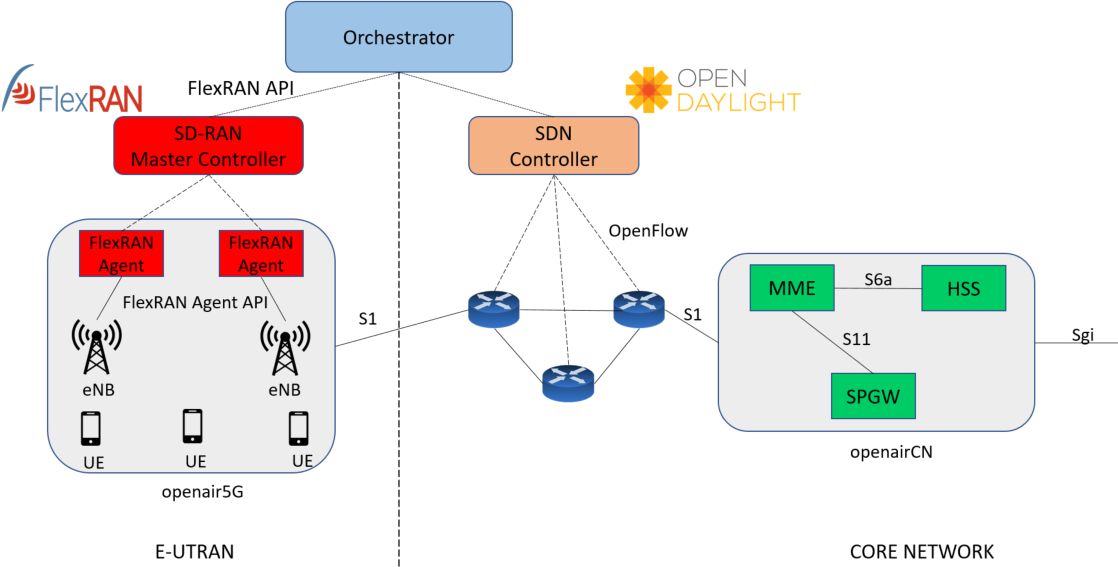}
\vspace{-0.2cm}
\caption{IqInf testbed architecture~\cite{inbook}.}
\vspace{-0.3cm}\label{fig:IqInf}
\end{figure}

\begin{figure}[h]
\centering
\includegraphics[scale=0.72]{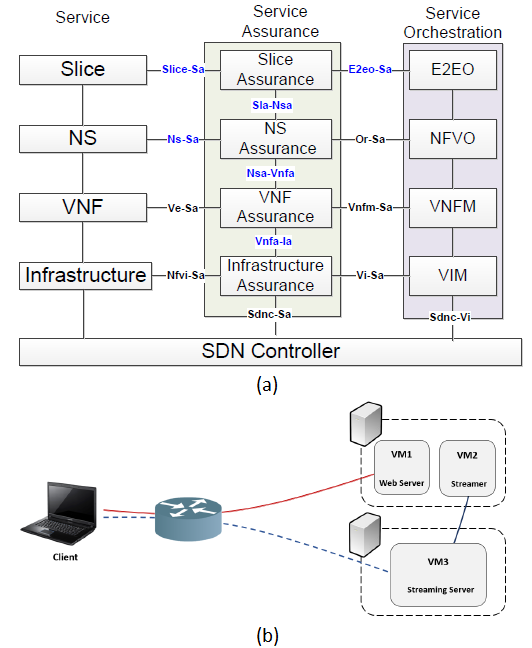}
\vspace{-0.1cm}
\caption{Slice-Aware SA framework (a). Architecture. (b). Measurement setup.~\cite{8806679}.}
\vspace{-0.3cm}\label{fig:Slice_Aware_SA_testbed}
\end{figure}

\begin{figure}[h]
\centering
\includegraphics[scale=0.9]{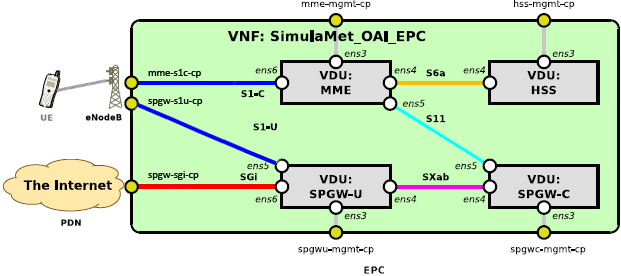}
\vspace{-0.5cm}
\caption{Simula testbed architecture~\cite{10.1007/978-3-030-44038-1_105}.}
\vspace{-0.2cm}\label{fig:Simula_testbed}
\end{figure}

\begin{figure}[h]
\centering
\includegraphics[scale=0.43]{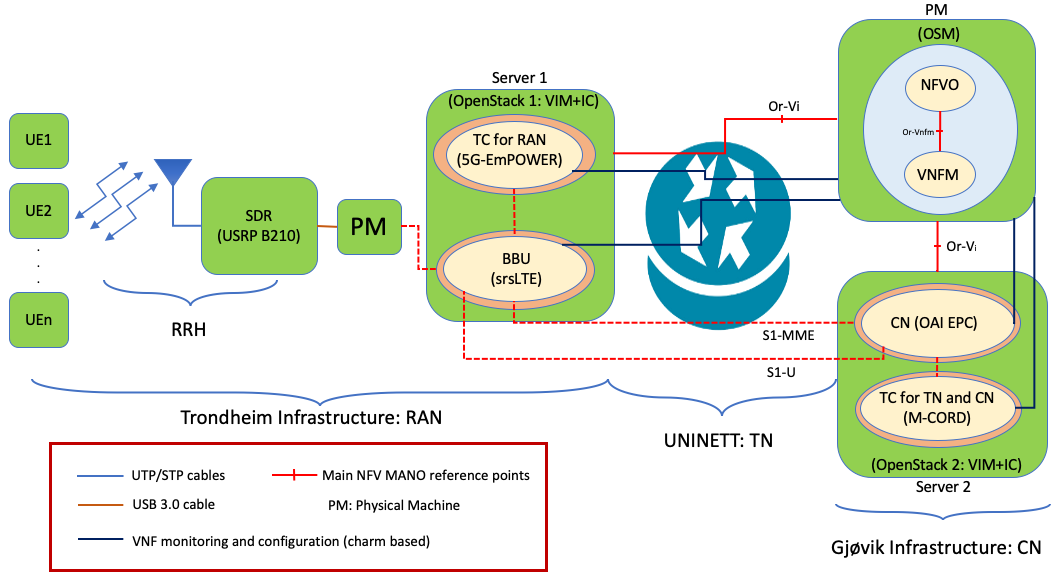}
\vspace{0cm}
\caption{5GIIK testbed architecture~\cite{9165419}.}
\vspace{-0.3cm}\label{fig:5GIIK}
\end{figure}

\begin{figure}[h]
\centering
\includegraphics[scale=1]{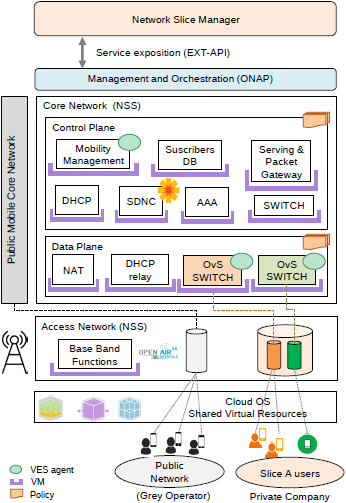}
\vspace{-0.1cm}
\caption{Architecture of the ONAP testbed integrated with slice management~\cite{9059507}.}
\vspace{-0.3cm}\label{fig:ONAP_based_testbed}
\end{figure}

\begin{figure}[h!]
\centering
\includegraphics[scale=0.505]{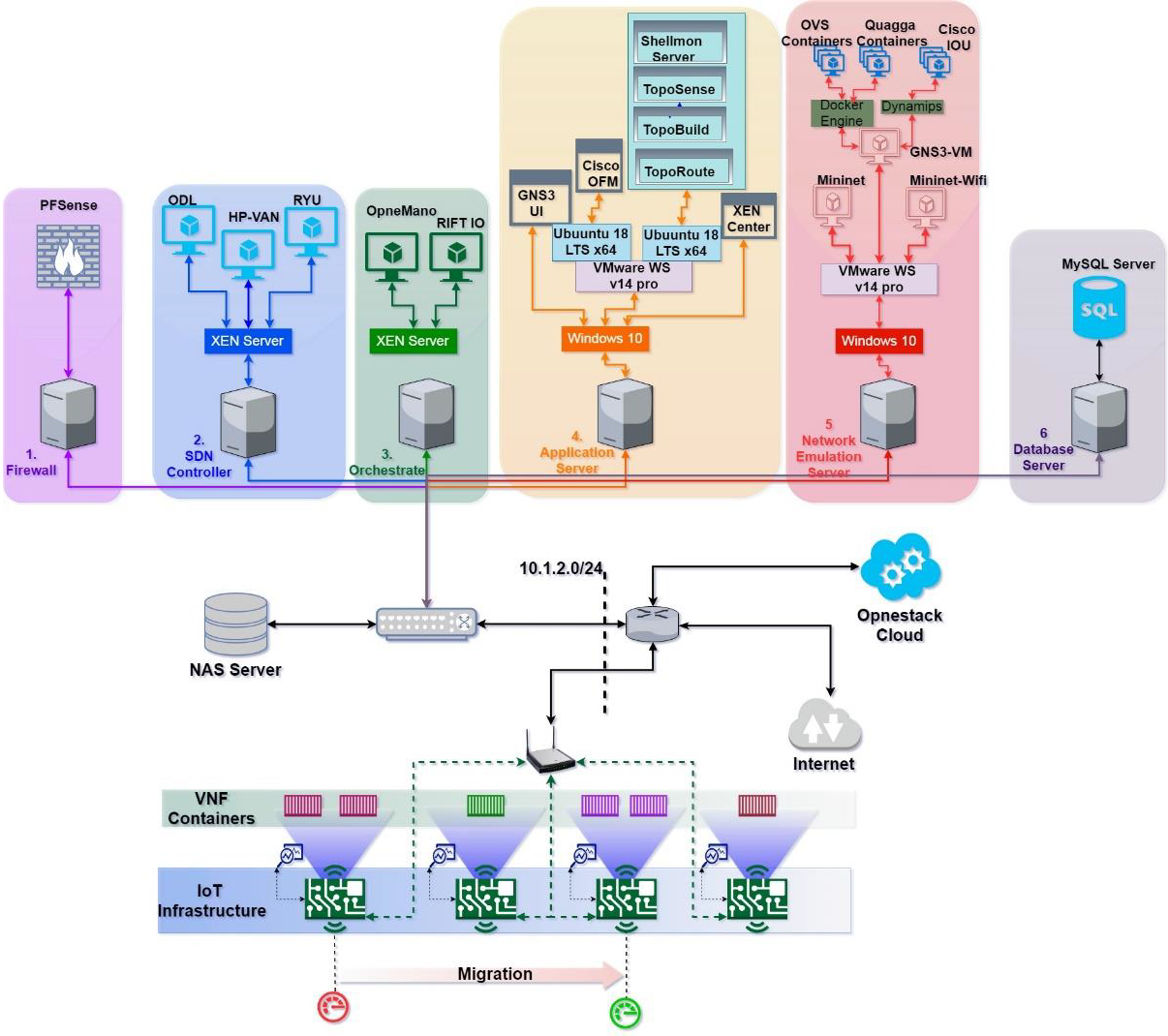}
\vspace{-0.5cm}
\caption{BlueArch testbed architecture~\cite{articlebluearch}.}
\vspace{-0.3cm}\label{fig:BlueArch}
\end{figure}

As an extension to the Clayman's model, here, slices are created via so-called transformable resources, which are interpreted as physically isolated (bare metal) or virtually shared resources. The VIMs are responsible for controlling and managing the number of allocated resources to each slice. The testbed utilizes the DC slice controller to deploy VIMs according to general templates for each slice dynamically. As a result, selecting a specific VIM converts to be a choice for a tenant and not a monopolized feature assigned by the network provider anymore, and it can be set based on the service specification. Hence, each tenant can now operate its own VIM. The testbed comprises a server running as the DC slice controller and four nodes with the same hardware configuration and equipped with an Arduino to trigger on/off action for each node and inspect each node's status. The testbed offers an evaluation scenario to determine the required time (loading, booting, configuration, and service startup times) to establish different infrastructures (VLSP, Kubernetes, and OpenStack).
\vspace{-0.2cm}
\\\\
\textit{4.12. Dynamic Slice Allocation Framework (DSAF)~\cite{Sattar2019DSAFDS}}. This testbed (Figure \ref{fig:DSAF}) is a practical implementation to evaluate the DSAF paradigm. Basically, DSAF is an efficient resource usage model for dynamic and real-time slice (de)allocation in the CN domain, which is based on minimum CPU utilization and finding links with the lowest delay. DSAF considers allocation policies for slice requests. DSAF also brings isolation between chained NFs of a slice. It is composed of five entities: 1) Orchestrator, which manages slice (de)allocation mechanism and all of the framework elements; 2) Optimization module, which monitors the available CPU and link delays while receiving slice requests; 3) Database, which maintains slice request information, slice allocation policies, and the available resources; 4) Optimization Agent, which acts as a mediator entity between the Orchestrator and the Optimization module to exchange information regarding slice allocation approaches; and 5) Hypervisor Agent, which interacts with the Orchestrator by presenting slice state information and performs slice (de)allocation. DSAF performance has been compared with First Come First Serve First Available (FCFSFA) method for different number of VNFs of a slice hosted by a Hypervisor. In these scenarios, the total number of slice requests allocated in DSAF is greater or equal than the FCFSFA scheme.
\vspace{-0.2cm}
\\\\
\textit{4.13. SliceNet platform~\cite{8667001}}. This testbed (Figure \ref{fig:SliceNet}) proposes a QoS-aware network slicing for multiple services with distinct QoS requirements. The testbed focuses on studying use cases for providing critical services with various reliability requirements. It introduces a novel SliceNet platform strategy to provide eHealth services via 5G network slicing. SliceNet offers a realistic network slicing with guaranteed QoS requirements by QoS-programmable policies in the data plane. This is done by implementing traffic engineering functions in both hardware and software levels. Moreover, SliceNet presents a plug and play control layer to let users demand customizable network slices in the network. SliceNet suggests E2E network slicing in both single and multi-domain providers. It also contributes to a cognitive network slice management functionality to enhance the QoS requirement for the services granted by the network slices. In addition to that, SliceNet also operates an ML-enabled method for the patient's examination in critical use cases via real-time video streaming communication from an ambulance to the medical center. This testbed is an extended version of the Mosaic5G platform~\cite{102623}.
\vspace{-0.2cm}
\\\\
\textit{4.14. Iquadrat Informatica (IqInf) testbed~\cite{inbook}}. This framework (Figure \ref{fig:IqInf}) utilizes OAI for deploying RAN and CN domains, and SDN switches (composed of Open vSwitch and VLAN switch) for building TN. The separation between CP and DP in the RAN domain is achieved by implementing FlexRAN Agent API, which provides a centralized load balancing and handover mechanism while having more than one eNB in this network. The OpenDayLight (ODL) realizes SDN policy in the TN domain. With the help of PHY abstraction mode of oaisim in OAI RAN, emulating practical network scenarios with numerous UEs and eNBs is conceivable. In particular, the OAI Traffic Generator (OTG) delivers network traffic of multiple applications like Voice over IP and Machine Type Communication (MTC) in this testbed. Deploying an orchestration scheme between SDN controllers and within the entire network domains has been considered as a future enhancement for the testbed architecture.
\vspace{-0.2cm}
\\\\
\textit{4.15. Slice-Aware Service Assurance (SA) Framework~\cite{8806679}}. This framework (Figure \ref{fig:Slice_Aware_SA_testbed}) examines Service Assurance (SA) in order to satisfy Quality of Experience (QoE) and QoS requirements in the context of network slicing. The framework integrates a novel SA-based architecture to the ETSI MANO platform to assure the services provided by different network slices in a network. Each component in the NFV MANO architecture has a counterpart in the SA-based NFV MANO platform: Slice Assurance, NS Assurance, NFV Assurance, and Infrastructure Assurance. These components operate four actions, including monitoring, analytics, management, and reporting, to guarantee the performance of the corresponding layer. This extended platform supports reporting information from all involved layers in service provisioning in the network slicing context. The platform also facilitates management and orchestration of various NFVs to assure that QoS and QoE requirements are fulfilled. The testbed evaluates the QoE of a service according to multiple service dependability Key Quality Indicators (KQIs). To this end, the testbed implements web content browsing and adaptive video streaming services to appraise infrastructure performance and the variation of KQIs for the service.
\vspace{-0.2cm}
\\
\textit{4.16. Simula Metropolitan Centre testbed~\cite{10.1007/978-3-030-44038-1_105, LCN2020-CloudRAN-Demo}}. This testbed (Figure \ref{fig:Simula_testbed}) demonstrates the deployment of OAI-EPC as a VNF on a cloud environment (OpenStack), and it presents the LTE CN service instantiation via OSM. In this testbed, according to the defined descriptors at the VNF and network service levels, the internal components of OAI-EPC are firstly cloned from related repositories. Secondly, they are implemented and configured via special configuration files, Juju Charms, on four separate virtual machines (Virtual Deployment Units (VDUs) in descriptors). The goals of this implementation are to produce MEC services to EPC as well as to integrate EPC with the extended eNB software. Finally, the testbed functionality is evaluated for establishing TCP and SCTP connections in three scenarios: downloading from server to UE, uploading from UE to the server, and bidirectional communication between UE and server. In its recent release, Simula testbed implements a mobile network based on OAI-EPC deployed as a VNF using OSM, which is now integrated with C-RAN architecture with functional split capability for BBU processing functions.
\vspace{-0.2cm}
\\\\
\textit{4.17. 5GIIK~\cite{9165419, DBLP:conf/nfvsdn/HagaEKG20}}. 5GIIK is a cross-location testbed (Figure \ref{fig:5GIIK}), which deploys OAI-EPC and srsLTE eNB as cloud-based VNFs via OSM on two OpenStack platforms launched at two geographically separated areas. In this testbed, the VNF-onboarding process takes place in three phases of the VNF lifecycle. In the first phase, management policies for establishing the VNFs are performed. In the second phase, configured VNFs grant the requested services. In the third phase, re-configuration of VNFs and monitoring of their Key Performance Indicators (KPIs) in runtime operation are provided. This testbed performs E2E network slicing via a hierarchical process by defining specific descriptors at the VNF, network service, and network slice levels on CN and RAN domains. The testbed also integrates 5G-EmPOWER for RAN, and M-CORD for TN and CN as SDN controllers. This results in supporting multi-tenancy in the RAN and also implementing slicing in the TN domains. It is worth noting that the 5G-EmPOWER assists common ML toolkits to facilitate realization and administration of machine learning models in this testbed. 5GIIK extends its capabilities by introducing Wireguard~\cite{wireguard:donenfeld} to its architecture as a Virtual Private Network (VPN)-based solution for providing slice isolation. In this solution, WireGuard-enabled VNFs operate on the NFVI via actions performed by OSM-NBI and Juju proxy charms. As a result, traffic isolation and security isolation, which are two essential features in network slicing, are granted via the integrated OSM-WireGuard framework.
\vspace{-0.2cm}
\\
\textit{4.18. Integrated Slice Management with ONAP framework~\cite{9059507}}. This testbed (Figure \ref{fig:ONAP_based_testbed}) investigates E2E network slicing lifecycle management (modeling, onboarding, instantiation, and operation) by integrating ONAP service orchestrator with a network slice manager entity. This integration grants a platform for 1) monitoring and collecting KPI reports that belong to the chained VNFs that create an E2E network slice and 2) evaluating the provided logs of information. In this way, multiple slices are inquired to trace whether the Service Level Agreement (SLA) between the service provider and service user is met or not. The testbed performs a use case by creating a private mobile network that affords services with best-effort and broadband QoS types via E2E network slices. Firstly, a slice is modeled as an ONAP-Network-Service composed of three VNFs (CP and DP for the CN domain, and a RAN emulator); besides, some policies for guaranteeing the SLA are defined. Secondly, the slice is deployed according to corresponding templates for each VNF. The testbed then utilizes the defined policies to perform slice management by modifying the allocated cloud resources to the two QoS types. Consequently, a dedicated channel grants a higher priority to broadband service compared to the best-effort service.
\vspace{-0.2cm}
\\\\
\textit{4.19. BlueArch~\cite{articlebluearch}}. BlueArch platform (Figure \ref{fig:BlueArch}) includes a customized structure of some open-source tools. The testbed serves in three operational modes: simulation, emulation, and access to a physical network to communicate with other platforms. The testbed architecture comprises two main sections. Section one consists of: 1) A Network Attached Storage (NAS) server representing shared storage that presents a private network; 2) A gateway router attaches a wireless access point operating on the same private network, an external OpenStack infrastructure, and the Internet; 3) Raspberry Pi accessories for MEC nodes implementation outcomes in producing IoT infrastructure in order to migrate VNFs. Section two consists of six VMs each encompassing a specific functionality: 1) An open-source PfSence~\footnote{\url{https://www.pfsense.org/}} firewall for conducting regular firewall actions and also for traffic shaping, network monitoring, and load balancing; 2) Employing ODL, Ryu, and HP-VAN SDN controllers hosted by a Citrix XEN server for yielding a cross-platform controlling of OvS devices in DP; 3) Open MANO and RIFT.io hosted by another XEN server that operate as orchestrators and support VNF-onboarding process for network slicing; 4) An application server works as the SDN application layer hosting open-source operating systems clients, which in turn driving GNS3 UI, a hypervisor, and XEN center; 5) A network emulation server involves two types of Mininet for wired and wireless SDN, and GNS3 Compute for offloaded computation resulting from GNS3 UI; 6) A MySQL-based database server interfacing the testbed with an external platform.

\begin{figure}[h]
\centering
\includegraphics[scale=0.58]{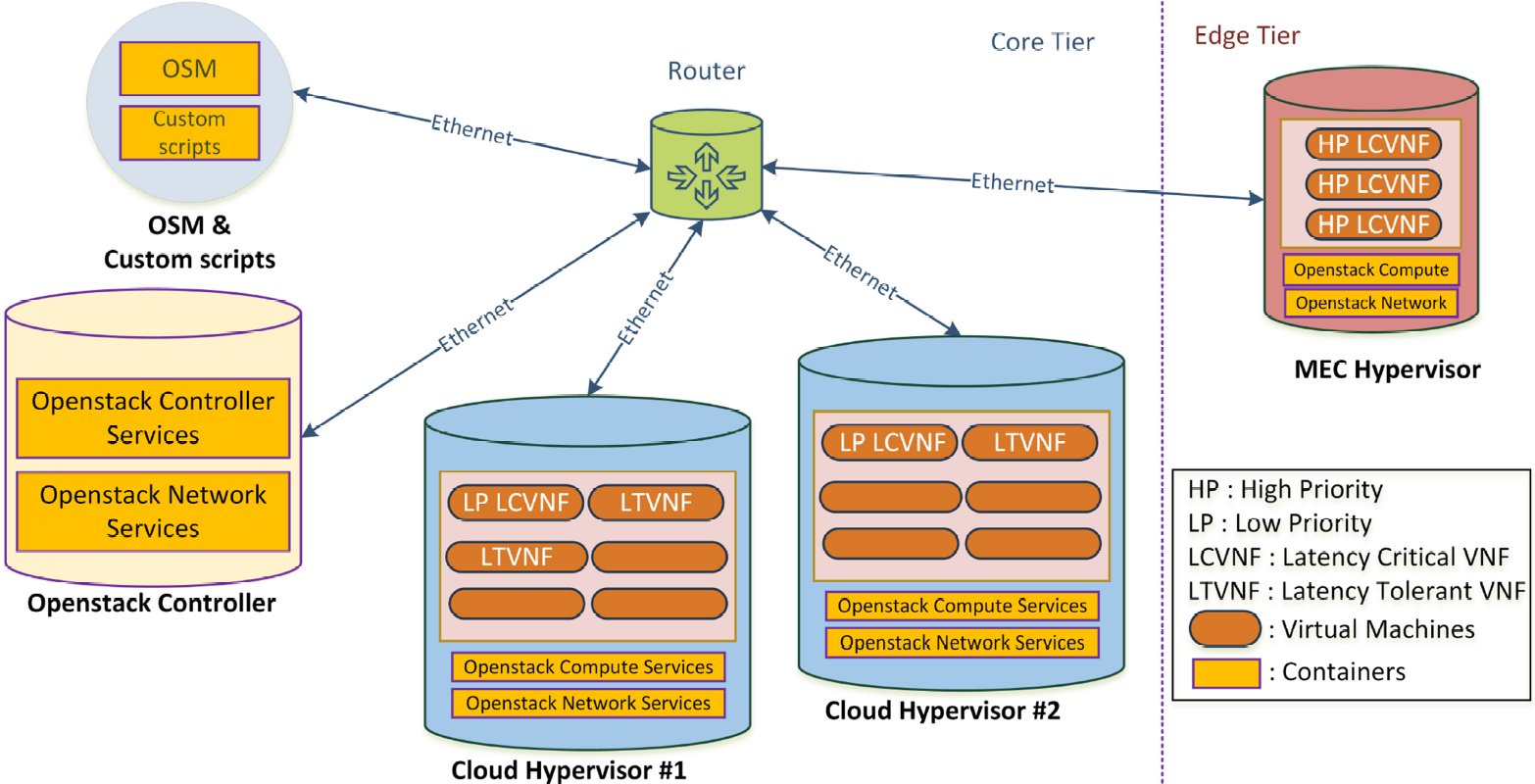}
\vspace{0cm}
\caption{MEC-Enabled 5G IoT architecture~\cite{8854289}.}
\vspace{-0.3cm}\label{fig:MEC_Enabled_5G_IoT_Architecture}
\end{figure}

\begin{figure}[h]
\centering
\includegraphics[scale=0.62]{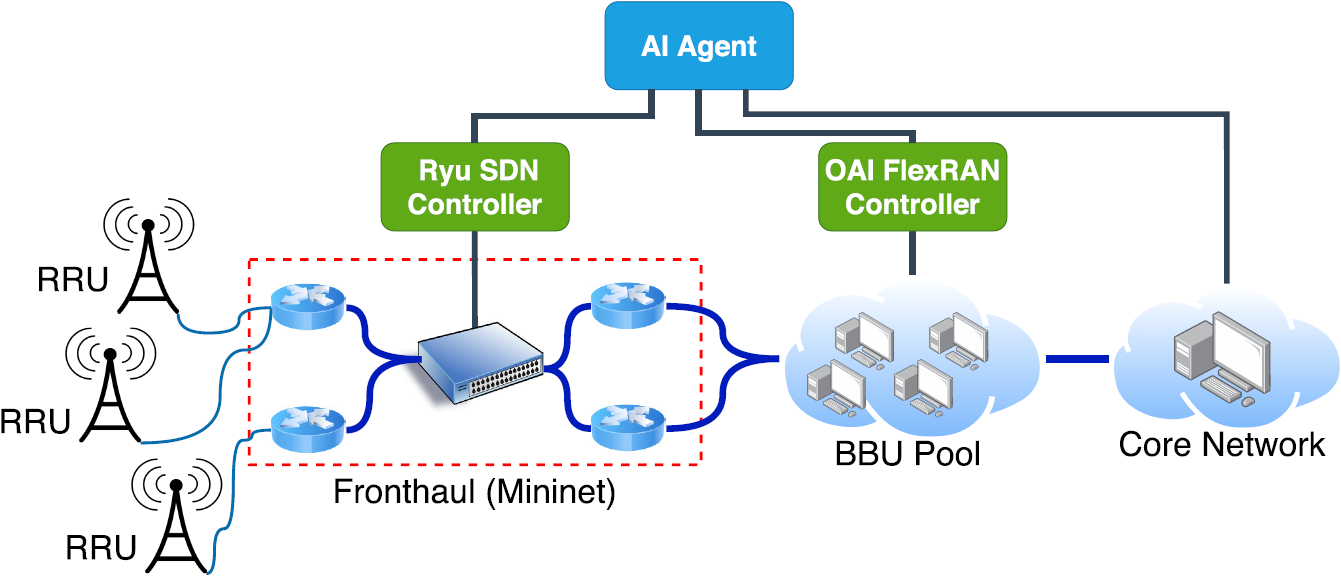}
\vspace{-0.3cm}
\caption{CAI testbed~\cite{9290141}.}
\vspace{-0.3cm}\label{fig:CAI}
\end{figure}

The testbed is evaluated in three use cases: 1) Real-time monitoring of resource utilization in disaster recovery by installing ShellMon client on IoT gateways; 2) Hosting VNF as a docker container when a MEC node becomes overloaded by taking a self-triggered action to relocate to another MEC node (known as VNF migration); 3) Modeling wireless channel and scheduling radio resources in RAN domain employing Matlab and using the testbed to perform SDN functionality.
\vspace{-0.2cm}
\\\\
\textit{4.20. MEC-Enabled 5G IoT platform~\cite{8514943, 8854289}}. This work (Figure \ref{fig:MEC_Enabled_5G_IoT_Architecture}) is a solid proposal for onboarding and scheduling aspects in VNF lifecycle management, and it presents a programmable and flexible MEC-enabled platform for IoT traffic. In this work, VNFs are categorized into Latency Critical VNFs (LCVNFs) and Latency Tolerant VNFs (LTVNFs). As a result, the applications are also divided into 1) real-time, provided by High Priority LCVNFs (HP LCVNFs), with resources in the MEC, 2) near-real-time, provided by Low Priority LCVNFs (LP LCVNFs), and 3) non-real-time, provided by LTVNFs. The LP LCVNFs and LTVNFs are deployed on the cloud instead of MEC since they do not provide real-time applications. The work improves the joint orchestration capability in the NFVO for the MEC and cloud resources for the mentioned VNF types via two methods: 1) an online placement scheme to deliver the required management tasks at the VNF level according to the data traffic, and 2) a latency embedding structure that enables VNF migration and scalability to fulfill service requirements in real-time. These two methods are accomplished by introducing 1) an algorithm for VNF Forwarding Graph (VNFFG) in chained VNFs for prioritizing delay-sensitive services, and 2) a second algorithm for the real-time allocation of the MEC and cloud resources to the VNFs that takes into account scale-in/out features for diverse service requirements. The testbed is deployed on several physical servers for the functionalities of the core (cloud infrastructure and NFVO) and network edge (MEC) with lower computational resources compared to the core. OpenStack, as the VIM with its telemetry feature, conducts data collection, data monitoring for future resource utilization, and placement policy through its compute schedulers. Furthermore, the OSM provides the NFVO functionality in this testbed. There are some hypervisors located at the core and the edge that afford the computing tasks. The testbed is assessed by some autoscaling, VNF placement, and online VNF scheduling scenarios.
\vspace{-0.2cm}
\\\\
\textit{4.21. CAI testbed~\cite{9290141}}. This testbed (Figure \ref{fig:CAI}) 
offers a cost-efficient virtualized and orchestrated 5G mobile network equipped with containers and distinct fronthaul and backhaul topologies. The testbed mainly concentrates on integrating Artificial Intelligence (AI) using Kubeflow tool~\cite{Kubeflow} to the management tasks in the 5G RAN and TN domains in order to optimize network performance. The testbed, called Connected AI (CAI), with the help of Kubernetes as a container-orchestrator, presents a mobile network composed of OvS devices, Ryu as SDN controller, and the OAI FlexRAN controller. CAI expedites the deployment of various network topologies on the fronthaul and backhaul by creating an emulated TN using Mininet. An AI agent takes various actions in the network by employing the information granted via Ryu and OAI FlexRAN controllers to feed ML models in order to implement several slice configurations. CAI builds a containerized implementation of OAI for C-RAN and Free5GC for the CN using Docker. The CAI testbed is evaluated via two use cases 1) monitoring the amount of allocated radio resource blocks to different slice requests and 2) VNF placement in a cluster of containers by means of the AI agent. 
\vspace{-0.2cm}

\section{5. Discussion}
\label{sec:Discussion and challenges}
\noindent
\textit{5.1. Comparison between different state-of-the-art network slicing testbeds}.
\label{sec:Comparison}
In this section, we compare the testbeds according to the design criteria for network slicing testbeds presented in Section 3. Table \ref{TableOverview} summarizes the major characteristics of each testbed. The testbeds in Table \ref{TableOverview} can be arranged into two categories. 
\begin{enumerate}[label={(\roman*)}]
\item The first category comprises those testbeds that partially achieve some of the \textit{primary} or \textit{secondary} attributes of the design criteria for network slicing testbeds. In this regard, the testbeds in~\cite{8405499, 8902745, Foukas2017OrionRS, foukas2018experience, articleUPC, 8627115, Sattar2019DSAFDS, inbook, 8806679, 10.1007/978-3-030-44038-1_105, articlebluearch, 8459990, 8514943, 8854289, 9290141} present network slicing in a particular network domain, and they do not realize a complete E2E network slicing. Reference~\cite{8459990} applies network slicing within multiple-VIMs (DCs); however, this implementation is limited to one network domain, and it does not present E2E network slicing, which crosses all network domains (RAN, TN, and CN). Other testbeds such as~\cite{865686122356} implements E2E network slicing; however, it does not offer MANO capability, multi-RATs, and multi-tenancy facilities in the architecture. The platform in~\cite{8718538} applies light employment of MANO entity and E2E network slicing in its design. 
\item The second category encompasses the implementations which satisfy all of the \textit{primary} and the majority of \textit{secondary} 
attributes from the design criteria explained in Section 3. The testbeds such as those in~\cite{102623,8524891, garcia2020experimenting, 8631816MCORD, 840611333MCORD, 9059507, 8667001, 9165419} deliver E2E network slicing with MANO privilege in their architectures along with multi-tenancy and multi-RATs support. The testbeds in~\cite{8667001, 9165419} also incorporate ML-enabled capability in their architectures, and the testbed in~\cite{9165419} is open-source.
\end{enumerate}
\begin{table}[p!]
\vspace{-0.1cm}
\caption{Comparison of small-scale testbeds for network slicing in 5G, where \checkmark denotes supported feature and \ding{55} denotes unsupported feature.}
\vspace{-0.4cm}
\begin{center}

\begin{minipage}{\textwidth} \centering

\begin{tabular}{ | m{2cm} | m{0.45cm}| m{0.4cm}| m{0.6cm} | m{0.75cm}| m{0.75cm}| m{0.75cm}| m{0.6cm} | m{0.6cm} | m{0.8cm} |m{0.8cm} | m{0.6cm} | m{1.5cm} | } 
  \hline
  \textbf{\scriptsize Testbed} & \textbf{\scriptsize SDN} & \textbf{\scriptsize NFV} & \textbf{\scriptsize Cloud comp.} & \textbf{\scriptsize Multi-domain} & \textbf{\scriptsize Multi-tenancy} & \textbf{\scriptsize MANO} & \textbf{\scriptsize Multi-RATs} & \textbf{\scriptsize E2E slicing} & \textbf{\scriptsize Cross-location} & \textbf{\scriptsize ML-enabled} & \textbf{\scriptsize Open-source} & \textbf{\scriptsize MANO type}\\ 
  \hline
  \scriptsize 1. 5G4IoT~\cite{8405499, 8854015}& \checkmark & \checkmark & \checkmark & \checkmark & \checkmark & \ding{55} & \ding{55} & \ding{55} & \ding{55}  & \ding{55} & \ding{55} & \ding{55} \\ 
\hline
  \scriptsize 2. 5GTN~\cite{8902745}& \checkmark & \checkmark & \checkmark & \checkmark & \ding{55} & \checkmark & \checkmark & \ding{55} & \ding{55} & \ding{55}  & \checkmark~\footnote{ \url{https://5gtn.fi/}} & \scriptsize OSM, CloudBand\\ 
\hline
  \scriptsize 3. SEMIoTICS~\cite{8718538}& \checkmark & \checkmark & \checkmark  & \checkmark & \checkmark & \checkmark & \ding{55} & \checkmark &\ding{55} &\ding{55} &\checkmark~\footnote{ \url{https://www.semiotics-project.eu/}} & \scriptsize OpenStack Tacker\\ 
  \hline
  \scriptsize 4. Mosaic5G~\cite{102623}& \checkmark & \checkmark & \checkmark  & \checkmark & \checkmark & \checkmark & \checkmark & \checkmark & \ding{55} & \ding{55}  &\checkmark~\footnote{ \url{http://mosaic5g.io/}}  &\scriptsize JOX\\ 
  \hline
  \scriptsize 5. Orion~\cite{Foukas2017OrionRS, foukas2018experience}& \checkmark & \checkmark & \checkmark & \checkmark & \checkmark & \checkmark & \ding{55} & \ding{55} & \checkmark & \checkmark  &\ding{55}  &\scriptsize OSM and a customized orchestrator\\ 
  \hline
  \scriptsize 6. 5G Testbed for NS~\cite{865686122356}& \ding{55} & \checkmark &  \checkmark & \checkmark & \ding{55} & \ding{55} & \ding{55} & \checkmark & \ding{55} & \ding{55} &\checkmark~\footnote{ \url{https://github.com/ashxz47}} & \ding{55}\\
  \hline
 \scriptsize 7. POSENS~\cite{8524891, garcia2020experimenting}& \checkmark & \checkmark & \checkmark  & \checkmark & \checkmark & \checkmark & \checkmark & \checkmark & \ding{55} & \ding{55} &\checkmark~\footnote{ \url{https://github.com/wnlUc3m}} &\scriptsize Customized OSM \\ 
  \hline
  \scriptsize 8. UPC testbed~\cite{articleUPC}& \checkmark & \checkmark & \ding{55} & \checkmark & \checkmark & \ding{55} & \checkmark & \ding{55} & \ding{55} & \checkmark &\ding{55} & \ding{55} \\ 
  \hline
  \scriptsize9. M-CORD based testbed~\cite{8631816MCORD, 840611333MCORD}& \checkmark & \checkmark & \checkmark & \checkmark & \checkmark & \checkmark & \checkmark & \checkmark & \ding{55} & \ding{55} &\checkmark~\footnote{ \url{https://nick133371.github.io/}} &\scriptsize XOS \\ 
  \hline
  \scriptsize 10. NS for 5G IoT and eMBB~\cite{8627115}& \checkmark & \checkmark & \checkmark  & \checkmark & \checkmark & \ding{55} & \checkmark & \ding{55} & \ding{55} & \ding{55} &\ding{55} & \ding{55}\\ 
  \hline
  \scriptsize 11. Transformable resources slicing testbed~\cite{8459990}& \checkmark & \checkmark & \checkmark & \ding{55}~\textsuperscript{*}& \checkmark & \checkmark & \ding{55} & \ding{55}~\textsuperscript{*}& \ding{55}~\textsuperscript{*}& \ding{55} & \ding{55} &\scriptsize VLSP, Kubernetes, and OpenStack \\ 
  \hline 
  \scriptsize 12. DSAF~\cite{Sattar2019DSAFDS}& \ding{55} & \checkmark & \checkmark & \ding{55} & \ding{55} & \checkmark & \ding{55} & \ding{55} & \ding{55} & \ding{55} &\ding{55} & \scriptsize Customized Python-based orchestrator \\ 
  \hline
  \scriptsize 13. SliceNet~\cite{8667001}& \checkmark & \checkmark & \checkmark  & \checkmark & \checkmark & \checkmark & \checkmark & \checkmark & \ding{55} & \checkmark & \ding{55} &\scriptsize OSM, OpenBaton \\ 
  \hline
  \scriptsize 14. IqInf testbed~\cite{inbook}& \checkmark & \checkmark & \ding{55} & \checkmark & \ding{55} & \ding{55} & \checkmark & \ding{55} & \ding{55}& \ding{55} &\ding{55} & \ding{55} \\
 \hline
 \scriptsize 15. Slice-Aware SA testbed ~\cite{8806679} & \checkmark & \checkmark & \ding{55} & \ding{55} & \ding{55} & \checkmark & \ding{55} & \ding{55} & \ding{55} & \ding{55} &\ding{55} &\scriptsize Service Assurance integrated with MANO\\ 
  \hline
  \scriptsize 16. Simula~\cite{10.1007/978-3-030-44038-1_105, LCN2020-CloudRAN-Demo} & \checkmark & \checkmark & \checkmark & \checkmark & \checkmark & \checkmark & \checkmark & \ding{55} & \ding{55} & \ding{55} &\checkmark~\footnote{ \url{https://github.com/simula/5gvinni-oai-ns}} &\scriptsize OSM \\ 
  \hline
 \scriptsize 17. 5GIIK~\cite{9165419, DBLP:conf/nfvsdn/HagaEKG20} & \checkmark & \checkmark & \checkmark & \checkmark & \checkmark & \checkmark & \checkmark & \checkmark & \checkmark&\checkmark &\checkmark~\footnote{\url{https://bit.ly/3rgOgd6}} &\scriptsize OSM \\
  \hline
  \scriptsize 18. ONAP based testbed~\cite{9059507} & \checkmark & \checkmark & \checkmark & \checkmark & \checkmark & \checkmark & \checkmark & \checkmark & \ding{55} & \ding{55} & \ding{55} &\scriptsize ONAP \\
  \hline
  \scriptsize 19. BlueArch~\cite{articlebluearch} & \checkmark & \checkmark & \checkmark & \checkmark & \ding{55} & \checkmark & \ding{55} & \ding{55} & \ding{55} & \ding{55} & \ding{55} &\scriptsize Open MANO, RIFT.io \\
  \hline
   \scriptsize 20. MEC IoT platform~\cite{8514943,8854289} & \checkmark & \checkmark & \checkmark & \checkmark & \ding{55} & \checkmark & \checkmark & \ding{55} & \ding{55} & \ding{55} & \ding{55} & \scriptsize OSM \\
   \hline
   \scriptsize 21. CAI~\cite{9290141} & \checkmark & \checkmark & \checkmark & \checkmark & \ding{55} & \ding{55} & \checkmark & \ding{55} & \ding{55} & \checkmark &\checkmark~\footnote{ \url{https://bit.ly/3tXErSX}} & \ding{55} \\
  \hline
\end{tabular}

\end{minipage}

\end{center}

\label{TableOverview}
\vspace{-0.4cm}{\textsuperscript{*}\footnotesize{E2E slice traverses over RAN, TN and CN.}}
\end{table}
\textit{5.2. Implementation challenges for deploying network slicing testbeds}.
\label{sec:Challenges}
This section presents some of the current challenges for deploying small-scale network slicing testbeds and summarizes proposed solutions that can slightly mitigate these challenges. 
\begin{enumerate}[label={(\roman*)}]
\item \textit{Monitoring frameworks for testbeds.} 5G is expected to provide heterogeneous services with distinct QoS requirements via utilizing network slicing. In this regard, the dynamic monitoring of the launched services is essential. This becomes challenging when recognizing the issues of possible performance degradation of the services. In fact, the multi-layered architecture of the 5G network, as shown in Figure \ref{fig:NFV_composition}, causes such challenges. Intelligently identifying such issues requires analyzing multiple possible sources of the problem via particular frameworks to effectively monitor the deployment and performance of services. To partially address this problem, different types of monitoring capabilities are integrated in some of the elaborated testbeds. The testbeds in which OSM acts as an orchestrator in their architectures, such as~\cite{8902745, Foukas2017OrionRS, foukas2018experience, 8667001, 10.1007/978-3-030-44038-1_105, 9165419, 8854289}, usually employ the interaction of the system Monitoring module (MON) with a monitoring toolkit such as Prometheus~\cite{Prometheus-website} for collection of VNFs' metrics and then utilize Grafana~\cite{Grafana-website} to visualize the collected data. The testbeds with an ONAP orchestrator, such as~\cite{9059507}, focus on SLA monitoring by exploiting Data Collection Analytics \& Events (DCAE) and Virtual Event Streaming (VES) components. Reference~\cite{102623} benefits from a monitoring application in the Store component. The architecture in~\cite{8806679} offers monitoring functions in each layer of SA and also implements virtual monitoring agents or virtual probes at each point of presence to actively observe network services.
\item \textit{Cross-location testbeds.} Launching testbeds over separate areas impacts the service performance because of delay, jitter, and packet loss. This issue becomes even more challenging when providing delay-sensitive services. Consequently, discovering techniques to enhance service performance in cross-location deployment is exceptionally important. As mentioned in the Table \ref{TableOverview}, the testbeds in~\cite{foukas2018experience, 9165419} deploy a cross-location architecture for C-RAN (RAN and MEC) and CN on two separate cloud-based infrastructures. In these two testbeds, the MANO entity (OSM), with the help of an SDN-assist feature, partially considers this issue by implementing application-aware traffic flow strategies to mitigate the generated latency because of the cross-location architecture, which results in enhancing connection reliability~\cite{foukas2018experience, osmjuly2020}. 
\item \textit{C-RAN deployment on testbeds.} Implementing C-RAN architecture on a testbed using open-source software packages can be challenging since the interaction between BBU and RRHs entails extremely low latency. Some attempts, such as in~\cite{102623, foukas2018experience, LCN2020-CloudRAN-Demo} resolve this problem by deploying the BBU section with a combination of PNF and VNF. They split the protocol stack of BBU into two sections in their solution instead of launching the BBU completely in a cloud-based environment. In particular, the functionality of the PHY layer of the BBU is split into a lower-PHY as PNF (to run on a physical machine along with RRHs) and higher-PHY as VNF (to run on a cloud infrastructure). In this way, the communication between (lower-PHY layer of) BBU and RRHs fulfills the ultra-low delay requirement while keeping the benefit of the cloud-based implementation of (higher-PHY layer of) BBU.
\item \textit{Resource management on testbeds with limited infrastructure capacity.} Resource management is considered as another possible challenge while deploying testbeds on infrastructures with limited physical and/or virtual resources. Since diverse services demand various amounts of networking, computing, and storage resources, it is essential to identify optimized methods to allocate available resources to service instances. To deal with this issue, testbeds that adopt OpenStack as VIM in their infrastructures, such as references~\cite{8405499, 8854015, 8902745, 8718538, 10.1007/978-3-030-44038-1_105, 9165419, Foukas2017OrionRS, foukas2018experience}, can enable Telemetry Data Collection to gather event and data for utilization statistics of the infrastructure resources.
\item \textit{Slice isolation on testbeds.} The (intra/inter) slice isolation concept is a common concern while implementing network slicing, and it is not limited to research testbeds. It is worth stating that there are some endeavors to tackle the isolation issue. 
Testbeds, such as those in~\cite{102623, 8627115, 8667001, Foukas2017OrionRS, foukas2018experience}, which utilize FlexRAN in their architectures, present partial slice isolation in the RAN domain. The testbeds in~\cite{8524891, garcia2020experimenting} perform isolation in the RAN domain by slicing the protocol stack down to RRC, RLC, and MAC layers. Nevertheless, introducing and realizing efficient and practical techniques to guarantee isolation in network slicing, especially in the RAN domain, is subject of future work. The work presented in~\cite{DBLP:conf/nfvsdn/HagaEKG20} is one step towards providing traffic isolation and security isolation in network slicing.
\end{enumerate}

\section{6. Conclusion}
\label{sec:Conclusion}
Network slicing testbeds with dedicated management and orchestration entities endeavor to outline and emulate trial and real use cases to achieve network slicing. On this basis and according to pioneer technologies, this paper addresses the principal design criteria for creating and deploying experimental environments for network slicing in 5G. After that, the paper explains the most common small-scale state-of-the-art testbeds for network slicing with their characteristics. The presented testbeds are then reviewed and compared via the design criteria, followed by possible challenges while creating such experimental platforms. Although many efforts have been performed to create testbeds for examining and evaluating network performance under various use cases in network slicing, there are still open research questions in this field. 
\section{Conflicts of Interest}~\label{sec:conflicts}
The authors declare that there is no conflict of interest regarding the publication of this paper.
\printbibliography
\end{document}